\documentclass[review]{elsarticle}

\usepackage{lineno}

\makeatletter
\def\ps@pprintTitle{%
 \let\@oddhead\@empty
 \let\@evenhead\@empty
 \def\@oddfoot{}%
 \let\@evenfoot\@oddfoot}
\makeatother
 \usepackage{threeparttable}
\usepackage{subfigure}
\usepackage{graphicx}
\usepackage{amsmath}
\usepackage{amssymb}
\usepackage{amsfonts}
\usepackage{mathtools}

\usepackage{stfloats}
\usepackage{float}
\usepackage{lipsum}
\usepackage{multirow}
\usepackage[pdftex,colorlinks,bookmarksopen,bookmarksnumbered,citecolor=red,urlcolor=red]{hyperref}

\usepackage{color} 

\restylefloat{figure}

\usepackage{enumitem}
\usepackage{algorithm}
\usepackage{algorithmic}

\journal{Reliability Engineering \& System Safety}

\begin{document}

\begin{frontmatter}

\title{\Large{Cross-Entropy Based Importance Sampling for Stochastic Simulation Models}}

\author[mymainaddress]{Quoc~Dung~Cao}

\author[mymainaddress]{Youngjun Choe\corref{mycorrespondingauthor}}
\cortext[mycorrespondingauthor]{Corresponding author}
\ead{ychoe@uw.edu}

\address[mymainaddress]{Department of Industrial and Systems Engineering,\\ University of Washington, Seattle, WA 98195, USA}



\begin{abstract}
To efficiently evaluate system reliability based on Monte Carlo simulation, importance sampling is used widely. The optimal importance sampling density was derived in 1950s for the deterministic simulation model, which maps an input to an output deterministically, and is approximated in practice using various methods. For the stochastic simulation model whose output is random given an input, the optimal importance sampling density was derived only recently. In the existing literature, metamodel-based approaches have been used to approximate this optimal density. However, building a satisfactory metamodel is often difficult or time-consuming in practice. This paper proposes a cross-entropy based method, which is automatic and does not require specific domain knowledge. The proposed method uses an expectation--maximization algorithm to guide the choice of a mixture distribution model for approximating the optimal density. 
The method iteratively updates the approximated density to minimize its estimated discrepancy, measured by estimated cross-entropy, from the optimal density. The mixture model's complexity is controlled using the cross-entropy information criterion. The method is empirically validated using {\color{black}extensive numerical studies} and applied to a case study of evaluating the reliability of wind turbine using a stochastic simulation model. 
\end{abstract}


\end{frontmatter}

\normalsize
\section{Introduction}
\textcolor{black}{Many computer simulation models can be generally categorized into deterministic simulation models and 
stochastic simulation models.} 
In a deterministic simulation model, the simulator will produce a fixed output for the same input values. In this paper, we focus on stochastic simulation models. Due to extra stochastic elements in the simulation model, given a fixed input value, the output is a random variable representing the outcome of the simulation. Despite its flexible modeling capacity, stochastic simulation models can be more complex and computationally costly to generate a simulation outcome. 

One of the applications of Monte Carlo simulation is to compute an estimator of a system's expected output. In Monte Carlo simulation, one draws inputs from a suitable probability distribution, calculates the output, and repeats the process multiple times to obtain different outputs. The estimate of the expected output can be the average of those simulated outputs. Monte Carlo simulation is widely used with both deterministic simulation models and stochastic simulation models. As the computing power advances over the past decades, the reliance on Monte Carlo simulation has increased in various engineering disciplines. Particularly in reliability engineering, Monte Carlo simulation is widely used to study the behavior and reliability of a system, often through estimating its failure probability. 

However, the reliability evaluation based on Monte Carlo simulation remains challenging. The more closely the  simulation model mimics the real system, the more computationally expensive each simulation run is. Moreover, highly reliable systems such as wind turbines or nuclear reactors require many simulation replications to encounter a rare event of interest such as a system failure, which translates to further demand for computational resources \cite{BOURINET2016210}. These challenges are exacerbated with stochastic simulation models, which often have to be run multiple times at the same input value. These challenges highlight the importance of improving computational efficiency of the reliability study with stochastic simulation models.

Among techniques to speed up Monte Carlo simulation in rare event probability estimation{\color{black}, also known as variance reduction techniques,} importance sampling (IS) is one of the most promising methods \cite{kroese:2011handbook,CADINI2014109}, especially with deterministic simulation models \cite{kahn1953}. IS can boost the simulation efficiency by reducing the number of required simulation runs to achieve a target variance of failure probability estimator \cite{choe2018uncertainty, DAI201286}. IS methods are also used with stochastic simulation models in various applications, such as finance \cite{glasserman2005}, insurance \cite{asmussen2000}, reliability \cite{heidelberger1995, balesdent2016rare, GONG2018199}, communication networks \cite{chang1994}, and queueing operations \cite{sadowsky1991, blanchet2009, blanchet2014}.

{\color{black} Variance reduction techniques for stochastic \textit{black-box} simulation models are relatively few. Most of existing techniques `open up' the stochastic black-box to gain the simulation efficiency, for example, by exploiting promising sample paths of underlying stochastic processes or by identifying important conditional events (e.g., splitting \cite{glasserman1996splitting,l2006splitting, morio2010overview}, subset simulation \cite{au2001estimation,song2009subset}, and conditional Monte Carlo \cite{asmussen2006improved,chan2011rare}). The existing methods that do not open up the stochastic black-box generally rely on importance sampling \cite{Choe2015,chen2017oracle}. In theory, stratification method (or stratified sampling) alone can improve the efficiency, but it is still used in conjunction with importance sampling in practice for greater efficiency gain \cite{graf2018adaptive}. 
}

Theoretically, if the inputs are drawn from the optimal IS density, the variance of the probability estimator will be minimized and we can save the most computational cost. However, the theoretically optimal IS density is generally not implementable in practice and often necessitates some approximations such as a metamodel-based method \cite{Choe2015,ECHARD2013232,CADINI2015188} or the cross-entropy (CE) method \cite{rubinstein1999, CHOI2017999}. Metamodel-based IS aims to represent the original high-fidelity, but black-box, simulation model by a surrogate model of which outputs can be more efficiently computed. From the surrogate model, the optimal IS density can be approximated to reduce the variance of the failure probability estimator \cite{DUBOURG2013}. From another perspective, CE-based IS aims to find an IS density that is as close as possible to the theoretically optimal IS density, where closeness is measured by the Kullback-Leibler divergence \cite{rubinstein1999}. The CE method has been widely used for estimating the optimal IS density with deterministic simulation models. However, to the best of our knowledge, there is no literature exploring the CE method's application to stochastic simulation models. Thus, this paper proposes a novel method called the cross-entropy based stochastic importance sampling (CE-SIS) to approximate the optimal IS density for stochastic simulation models.  

{\color{black}The main contribution of the proposed CE-based method is the \textit{automation} of approximating the optimal IS density for \textit{stochastic} simulation models. In the literature, metamodel-based IS methods are predominant \cite{Choe2015,chen2017oracle}. The construction of a metamodel of a stochastic simulation model often requires substantial efforts of engineers and statisticians to appropriately model the additional randomness embedded in the simulation model. For example, the wind turbine simulation model in our case study uses over 8 million random variables to simulate 3-dimensional time-marching wind field in each run. It is challenging to build a metamodel to capture such large uncertainties based on a few hundreds of pilot runs of the simulation model, as detailed in \cite{Choe2015}. As demonstrated in our case study in Section~\ref{sec:case_study}, our proposed method achieves comparable estimation performance without such efforts and resources to build a highly sophisticated metamodel. Also, as discussed in our numerical studies in Section~\ref{sec:num_ex}, a poorly constructed metamodel can even substantially undermine the IS performance. In contrast, the proposed method performs reliably in a wide variety of settings (except for the cases where the user already knows the method should fail), showing the promise for various applications in practice.}  



One challenge of the CE-based IS is the dilemma between parametric and nonparametric representation of the candidate optimal IS density. In the standard CE method for deterministic simulation models, the candidate IS density is confined to a parametric family. Some of the popular parametric IS densities are multivariate Bernoulli and multivariate Gaussian, thanks to their simple random variate generation and convenient updating formulae for CE minimization. However, parametric IS can be too rigid to capture the complicated important region and it is often difficult to validate the parametric model assumptions \cite{botev2013}. Nonparametric approach can offer flexibility to overcome such limitations. One of the most popular nonparametric density approximation methods is the kernel approach \cite{MORIO2011178}. Besides its flexibility, the kernel approach also has its own drawback. The probability density model is not as parsimonious as (a mixture of) parametric density models. This lack of parsimony often makes subsequent inference and analysis of the resulting model computationally intensive \cite{rubinstein2005,botev2007}. For stochastic simulation models, the parametric IS approach needs a strong assumption but the variance of the IS estimator converges to the optimal variance faster than the nonparametric IS approach \cite{chen2017oracle}. The nonparametric IS approach requires weak assumptions but the variance reduction rate is not as fast as the parametric IS approach \cite{chen2017oracle}. To achieve a good balance between flexibility and efficiency in the CE-SIS method, we express the candidate IS density for stochastic simulation models using a mixture of parametric distributions. We particularly focus our study on the Gaussian mixture model due to its flexibility to model a smooth distribution. 

The proposed method is validated with {\color{black}extensive numerical studies} and a case study on wind turbine reliability evaluation. In the case study, our method is benchmarked against a state-of-the-art metamodel-based IS method in the literature. Approximating the optimal IS density automatically without relying on expert domain knowledge (which is often necessary for metamodel construction), the CE-SIS method demonstrates comparable results. This shows the advantage of the CE-SIS method in situations when building a high-quality metamodel is difficult or time-consuming; the CE-SIS method would conveniently  provide substantial computational saving potentially more than metamodel-based methods. 

\vspace{-.3cm}

\section{Background}\label{sec:background}
The crude Monte Carlo (CMC) method samples the input, $\mathbf{X} \in \mathbb{R}^p$, from a known probability density function, $f(\mathbf{x})$ \cite{kroese:2011handbook} . From the input $\mathbf{X}$, a simulator generates the output, $Y \in \mathbb{R}$.  If the simulation model is deterministic, $Y$ is a deterministic function of $\mathbf{X}$, i.e., $Y = g(\mathbf{X})$. For the stochastic simulation model, $Y$ is stochastic for a given $\mathbf{X}$. Due to the random vector $\boldsymbol{\epsilon}$ within the simulation model, it generates different outputs $Y$ even at a fixed input $\mathbf{X}$. 


In reliability engineering, a quantity of interest is the so-called failure probability, $\mathbb{P}\left( Y >l \right)$, where $l$ is a pre-specified threshold on a system that fails if $Y$ exceeds $l$. Note that any failure event set can be expressed as $\{Y >l\}$ using a transformation. To estimate the failure probability, the following CMC estimator is  most commonly used,{\color{black}
\begin{align}
  	\hat{P}_{CMC} &=	\frac{1}{n}\sum_{i=1}^{n}\mathbb{I}\left(Y_i > l\right)  ,	\label{eq:CMC_estimator}
\end{align}
where $n$ is the total number of simulation replications, and $\mathbb{I}\left(Y_i > l\right)$ is an indicator function that takes value of $1$ if $Y_i > l$ and value of $0$, otherwise}. The estimator in \eqref{eq:CMC_estimator} is an unbiased estimator for $\mathbb{P}\left( Y >l \right)$. For the deterministic simulation model, the failure probability is equivalent to $\mathbb{E}_f \!\left[ \mathbb{I}\left(g(\mathbf{X}) >l\right) \right]$, where $\mathbb{I}(\cdot)$ is the indicator function and the subscript, $f$, appended to the expectation operator, $\mathbb{E}$, denotes that the expectation is taken with respect to $f$, the density from which $\mathbf{X}$ is drawn. For the stochastic simulation model, the failure probability is the expectation of the conditional failure probability, expressed as $\mathbb{E}_f \!\left[ \mathbb{P}\left(Y >l \mid \mathbf{X}\right) \right]$.  

In a highly reliable system, the failure probability can be very low. It may take many simulation runs until a failure occurs. In some cases, each simulation run can be computationally very expensive. In order to obtain a good estimator of the failure probability, the number of simulation runs can be large. To save the computational resource, IS changes the sampling distribution of $\mathbf{X}$ from $f\!\left(\mathbf{x}\right)$ to another distribution $q\!\left(\mathbf{x}\right)$ that makes the failure events more likely. Sampling $\mathbf{X}$ from $q\!\left(\mathbf{x}\right)$ makes the estimator in \eqref{eq:CMC_estimator} no longer unbiased.

For the deterministic simulation model, to make the failure probability estimator unbiased, the IS estimator becomes
\begin{align}
\hat{P}_{DIS}	&=	\frac{1}{n}\sum_{i=1}^{n}   \mathbb{I}\left(Y_i > l\right)        \frac{f\!\left(\mathbf{X}_i\right)}{q\!\left(\mathbf{X}_i\right)}, 	 \label{eq:DIS_estimator}
\end{align}
where $\mathbf{X}_i$, $i = 1,\ldots,n$, is sampled from $q\!\left(\mathbf{x}\right)$ instead of $f\!\left(\mathbf{x}\right)$ \cite{kroese:2011handbook}. $Y_i$ is the output from the simulator corresponding to the input $\mathbf{X}_i$. The variance of $\hat{P}_{DIS}$ is minimized if $\mathbf{X}$ is sampled from the optimal IS density \cite{kahn1953}
\begin{align}
q_{DIS}\!\left(\mathbf{x}\right)&= \frac{ \mathbb{I}\left(g(\mathbf{x}) > l\right)  f\!\left(\mathbf{x}\right)}{\mathbb{P}\!\left(Y > l\right)	}.	 \label{eq:DIS_opt_density}
\end{align}

Here, the indicator function $\mathbb{I}\left(g(\mathbf{x}) > l\right)$ can be evaluated only by running the simulator since the function $g(\mathbf{x})$ is unknown. The denominator $\mathbb{P}\!\left(Y > l\right)$ is the unknown quantity that we want to estimate. Thus, $q_{DIS}$ is not implementable in practice, and approximation of $q_{DIS}$ is required. {\color{black}As mentioned above, for the deterministic simulation model, the optimal IS density can be approximated using methods such as the metamodel-based method (\textit{e.g.}, by building a metamodel of $g(\mathbf{x})$ or $\mathbb{I}\left(g(\mathbf{x}) > l\right)$ to construct a parametric or nonparametric IS density) and the CE method (\textit{e.g.}, by finding a parametric IS density close to $q_{DIS}$ in terms of CE).} Hereafter, we call the IS method for a deterministic simulation model DIS.

For the stochastic simulation model, to capture the extra randomness, the unbiased IS estimator becomes
\begin{align}
\hat{P}_{SIS}	&=	\frac{1}{m}\sum_{i=1}^{m} \left(    \frac{1}{N_i} \sum_{j=1}^{N_i} \mathbb{I}\left(Y_{j}^{(i)} > l\right)     \right)      \frac{f(\mathbf{X}_i)}{q(\mathbf{X}_i)} , 	\label{eq:SIS_estimator}
\end{align}
where $m$ is the number of distinct input $\mathbf{X}_i$ {\color{black}at which the stochastic simulation model is run $N_i$ times to obtain the outputs, $Y_{j}^{(i)}$, $j=1,\ldots,N_i$. Thus, $\sum_{i=1}^{m}N_i$ is equal to the total number of simulation replications, $n$}. Such $N_i$ replications at the same $\mathbf{X}_i$ value capture the additional randomness $\boldsymbol{\epsilon}$ within the simulation model. Hereafter, we call the IS method for a stochastic simulation model SIS.

Similar to DIS, the variance of $\hat{P}_{SIS}$ is desired to be as small as possible. Minimizing the variance requires two steps \cite{Choe2015}:

(a) sampling $\mathbf{X}$ from
\begin{align}
q_{SIS}\!\left(\mathbf{x}\right)	&= \frac{1}{C_{q}} f\!\left(\mathbf{x}\right)	\sqrt{	\frac{1}{n} s\!\left(\mathbf{x}\right)\left( 1- s\!\left(\mathbf{x}\right)\right)    +		 s\!\left(\mathbf{x}\right)^2 		}  ,   \label{eq:SIS_opt_density}
\end{align}
where $s\!\left(\mathbf{x}\right)$ is $\mathbb{P}\!\left(Y >l \mid \mathbf{X} = \mathbf{x}\right)$ and $C_q$ is the normalizing constant; and

(b) allocating replications to $\mathbf{X}_i$ by
\begin{align}
N_i &= n
\frac
{ 		 	\sqrt{    \frac{n \left(1-  s\left(\mathbf{X}_i\right) \right)  }{1 + \left(n-1\right)  s\left(\mathbf{X}_i\right) }     }    	   }
{ \sum_{j=1}^{m}   \sqrt{    \frac{n \left(1-  s\left(\mathbf{X}_j\right) \right)  }{1 + \left(n-1\right)  s\left(\mathbf{X}_j\right) }     }      }, \;\;\; i=1,\ldots,m. \label{eq:N_i_opt}
\end{align}

Since the conditional probability, $s\!\left(\mathbf{x}\right)$, is unknown in practice, the IS method for a stochastic simulation model requires the approximation of $s\!\left(\mathbf{x}\right)$, $q_{SIS}$, and $N_i$. This paper will focus on how the CE method can be extended to approximate $q_{SIS}$. In the next subsection, we first review how the CE method is applied to approximate $q_{DIS}$.

\subsection{CE Method for DIS}\label{sec:standardCE}

The CE method is originally developed to find the density that best approximates the optimal density of DIS \cite{rubinstein1999}. The standard parametric CE method limits the search space for the optimal IS density $q^*(\mathbf{x})$ to a pre-specified parametric family (e.g., Gaussian, Poisson, gamma, etc.), $\{q\!\left(\mathbf{x} ; \boldsymbol{\theta} \right)\!: \boldsymbol{\theta} \in \boldsymbol{\Theta}(d) \subset \mathbb{R}^d  \}$, and seeks the density $q\!\left(\mathbf{x} ; \boldsymbol{\theta}^* \right)$ that is closest to the optimal density $q^*(\mathbf{x})$. The closeness is measured by the Kullback-Leibler divergence \cite{rubinstein1999},
\begin{align}
\mathbb{D}\!\left(q^*(\mathbf{x}), q(\mathbf{x};\boldsymbol{\theta})\right)
&= \int  q^*\!\left(\mathbf{x}\right) \ln{q^*\!\left(\mathbf{x}\right)}\,\mathrm{d}\mathbf{x}
- \int  q^*\!\left(\mathbf{x} \right) \ln{q\!\left(\mathbf{x}; \boldsymbol{\theta}\right)}\,\mathrm{d}\mathbf{x} . \label{eq:KL_divergence}
\end{align}
This quantity is always non-negative and takes zero if and only if $q^*\!\left(\mathbf{x}\right) = q\!\left(\mathbf{x} ; \boldsymbol{\theta} \right)$ almost everywhere. Thus, minimizing $\mathbb{D}\!\left(q^*(\mathbf{x}), q(\mathbf{x};\boldsymbol{\theta})\right)$ over $\boldsymbol{\theta} \in \boldsymbol{\Theta}$ leads to $q\!\left(\mathbf{x} ; \boldsymbol{\theta}^*\right) = q^*\!\left(\mathbf{x}\right)$ if $q^*$ belongs to the same parametric family as $q\!\left(\mathbf{x} ; \boldsymbol{\theta}^*\right)$. 

Minimizing $\mathbb{D}\!\left(q^*(\mathbf{x}), q(\mathbf{x};\boldsymbol{\theta})\right)$ in \eqref{eq:KL_divergence} over $\boldsymbol{\theta}$ is equivalent to minimizing its second term, known as the cross-entropy (CE)
\begin{align}
\mathbb{C}(q^*(\mathbf{x}), q(\mathbf{x};\boldsymbol{\theta})) &= -\int q^*\!\left(\mathbf{x}\right) \log{q\left(\mathbf{x} ; \boldsymbol{\theta}\right)} \,\mathrm{d}\mathbf{x}, \label{eq:cross_entropy}
\end{align}
because the first term in \eqref{eq:KL_divergence} is a constant over $\boldsymbol{\theta}$. As shown in \eqref{eq:DIS_opt_density}, the optimal IS density can be expressed as $q^*\!\left(\mathbf{x}\right) \propto h(\mathbf{x})f(\mathbf{x})$, where $h(\mathbf{x})$ is $\mathbb{I}\left(g(\mathbf{x}) > l\right)$ for DIS. The CE method aims to equivalently minimize
\begin{align}
\mathcal{C}(\boldsymbol{\theta}) &= - \int h(\mathbf{x})f(\mathbf{x})  \log{q\left(\mathbf{x} ; \boldsymbol{\theta}\right)} \,\mathrm{d}\mathbf{x} \nonumber \\
&= - \mathbb{E}_f[h(\mathbf{X})  \log{q\left(\mathbf{X} ; \boldsymbol{\theta}\right)}] \nonumber \\
&= - \int h(\mathbf{x})\frac{f(\mathbf{x})}{q\left(\mathbf{x} ; \boldsymbol{\theta'}\right)}  \log{q\left(\mathbf{x} ; \boldsymbol{\theta}\right)} q\left(\mathbf{x} ; \boldsymbol{\theta'}\right)\,\mathrm{d}\mathbf{x} \label{eq:change_of_dist}\\
&= - \int h(\mathbf{x})w(\mathbf{x}; \boldsymbol{\theta}')  \log{q\left(\mathbf{x} ; \boldsymbol{\theta}\right)} q\left(\mathbf{x} ; \boldsymbol{\theta'}\right)\,\mathrm{d}\mathbf{x} \nonumber\\
&= - \mathbb{E}_q[h(\mathbf{X})w(\mathbf{X};\boldsymbol{\theta}')  \log{q\left(\mathbf{X} ; \boldsymbol{\theta}\right)}], \label{eq:CE_equiv}
\end{align}
where $\boldsymbol{\theta}$, $\boldsymbol{\theta}'$ $\in \boldsymbol{\Theta}(d)$, and $f(\mathbf{x})$ is the original density of $\mathbf{X}$. The likelihood ratio ${f(\mathbf{x})}/{q\left(\mathbf{x} ; \boldsymbol{\theta'}\right)}$ is denoted by $w(\mathbf{x};\mathbf{\boldsymbol\theta'})$. Note that the equality in \eqref{eq:change_of_dist} holds under the condition that $q\left(\mathbf{x} ; \boldsymbol{\theta'}\right) = 0$ implies $h(\mathbf{x})f(\mathbf{x}) = 0$. This condition can be easily satisfied by ensuring that the support of $q$ includes the support of $f$ even if $h$ is unknown. The condition is always satisfied by the Gaussian mixture model. In practice, the CE method finds $\boldsymbol{\hat{\theta}}$ that minimizes the following IS estimator of \eqref{eq:CE_equiv},
\begin{align}
\bar{\mathcal{C}}({\boldsymbol{\theta}}) &= - \frac{1}{n}\sum_{i=1}^{n} h(\mathbf{X}_i) w(\mathbf{X}_i) \log{q\!\left(\mathbf{X}_i ; \boldsymbol{\theta}\right),}  \label{eq:CE_estimator}
\end{align}
where $\mathbf{X}_i, i = 1,\ldots, n$, is sampled from $q(\mathbf{x}; \boldsymbol{\hat{\theta}}')$. Note that by writing $w(\mathbf{X}_i)$, we surpress the notation for dependence of $w$ on $\boldsymbol{\hat{\theta}}'$ in \eqref{eq:CE_estimator}. Using this IS estimator, the CE method minimizes the CE iteratively:
\begin{enumerate}[leftmargin=3.8em]
\item[Step 1.] Sample $\mathbf{X}_i, i = 1,\ldots, n$, from $q(\mathbf{x};  \boldsymbol{\hat{\theta}}')$. At the first iteration,  $q(\mathbf{x};  \boldsymbol{\hat{\theta}}')$ can be flexible (e.g., $f$ is commonly used).
\item[Step 2.] Find $\boldsymbol{\hat{\theta}} = \arg\!\min_{\boldsymbol{\theta}\in \boldsymbol{\Theta}} \bar{\mathcal{C}}({\boldsymbol{\theta}})$, where $\bar{\mathcal{C}}({\boldsymbol{\theta}})$ is in \eqref{eq:CE_estimator}.
\item[Step 3.] Set $\boldsymbol{\hat{\theta}}' = \boldsymbol{\hat{\theta}}$ and start the next iteration from Step 1 until some stopping criterion is met.
\end{enumerate}

This procedure iteratively refines $q(\mathbf{x} ; \boldsymbol{\hat{\theta}})$.  However, the refinement is limited, as the parameter search space, $\boldsymbol{\Theta}$, is restricted by a pre-defined parametric family. 

Some studies \cite{rubinstein2005,botev2007}  explore  nonparametric approaches to allow greater flexibility on the candidate IS density than the standard CE method. However, as mentioned before, the flexibility comes with costs: finding the optimal density \cite{botev2007} or sampling from the optimized density \cite{rubinstein2005} is computationally challenging. {\color{black}Essentially, the two extremes of the spectrum on expressing the candidate IS density depend on the relationship between $d$ (the number of parameters in a candidate IS density) and $n$ (the total number of simulation replications). A candidate IS density can be parametric with $d \ll n$ or nonparametric with $d \asymp n$ (\textit{i.e.}, $d$ has the same order of magnitude as $n$). To bridge the gap between the two extremes, recent studies} \cite{botev2013,Wang2015,kurtz2013,GEYER201915} consider the mixture of parametric distributions, where $d$ can vary between $1$ and $n$. This approach is particularly desirable for engineering applications because (a) it can be as flexible as we want; (b) it is easy and fast to sample from the mixture of parametric distributions; and (c) the mixture IS density provides an insight into the engineering system (e.g., means of mixture components often coincide with the so-called `hot spots', where the system likely fails.). Particularly in \citep{Choe2017}, the candidate distribution is expressed by the Gaussian mixture model (GMM) and an expectation--maximization (EM) algorithm is used to minimize the cross-entropy estimator in \eqref{eq:CE_estimator}. {\color{black}More recently and independently, another group of researchers \cite{GEYER201915} also proposes fundamentally the same EM algorithm to fit a GMM within the CE method.} 

{\color{black}
\subsection{Gaussian Mixture Model and EM Algorithm}\label{sec:GMM_EM}
The GMM of $k$ mixture components takes the following form:
\begin{align}
q\!\left(\mathbf{x} ; \boldsymbol{\theta}\right) &= \sum_{j=1}^{k} \alpha_j \, q_j\!\left(\mathbf{x} ; \boldsymbol{\mu}_j, \boldsymbol{\Sigma}_j\right) , \label{eq:GMM}
\end{align}
where the component weights, $\alpha_j$, $j= 1,\ldots,k$, are positive and sum to one. The $j$th Gaussian component density, $q_j$, is specified by the mean, $\boldsymbol{\mu}_j$, and the covariance $\boldsymbol{\Sigma}_j$. Thus, the parameter vector of GMM $\boldsymbol{\theta}$ denotes $\left(\alpha_1,\ldots,\alpha_k, \boldsymbol{\mu}_1, \ldots, \boldsymbol{\mu}_k, \boldsymbol{\Sigma}_1,\ldots, \boldsymbol{\Sigma}_k \right)$. Since $\boldsymbol{\theta}$ is a function of $k$, the GMM model can be explicitly written as $q\!\left(\mathbf{x} ; \boldsymbol{\theta}\right(k))$. For simplicity, we will write the GMM as $q\!\left(\mathbf{x} ; \boldsymbol{\theta}\right)$ unless we should express different models in terms of $k$.

To minimize \eqref{eq:CE_estimator}, the gradient of \eqref{eq:CE_estimator} with respect to $\boldsymbol{\theta}$ is set to zero:
\begin{align}
- \frac{1}{n}\sum_{i=1}^{n} h(\mathbf{X}_i) w(\mathbf{X}_i) \nabla_{\boldsymbol{\theta}} \log{q\left(\mathbf{X}_i ; \boldsymbol{\theta}\right)} &= 0 .
\end{align}
This leads to the updating equations as derived in \cite{kurtz2013}:
\begin{align}
\alpha_j &= \frac{\sum_{i=1}^{n} h(\mathbf{X}_i) w(\mathbf{X}_i) \gamma_{ij} }{\sum_{i=1}^{n} h(\mathbf{X}_i) w(\mathbf{X}_i)  } , \label{eq:alpha_updating}
\\
\boldsymbol{\mu}_j &= \frac{\sum_{i=1}^{n} h(\mathbf{X}_i) w(\mathbf{X}_i) \gamma_{ij} \mathbf{X}_i }{\sum_{i=1}^{n} h(\mathbf{X}_i) w(\mathbf{X}_i) \gamma_{ij} } , \label{eq:mu_updating}
\\
\boldsymbol{\Sigma}_j &= \frac{\sum_{i=1}^{n} h(\mathbf{X}_i) w(\mathbf{X}_i)  \gamma_{ij} (\mathbf{X}_i - \boldsymbol{\mu}_j)(\mathbf{X}_i - \boldsymbol{\mu}_j)^T}{\sum_{i=1}^{n} h(\mathbf{X}_i) w(\mathbf{X}_i) \gamma_{ij} } , \label{eq:sigma_updating}
\end{align}
where
\begin{align}
\gamma_{ij} &= \frac{ \alpha_j \, q_j\!\left(\mathbf{X}_i ; \boldsymbol{\mu}_j, \boldsymbol{\Sigma}_j\right)}{\sum_{j'=1}^{k} \alpha_{j'} \, q_{j'}\!\left(\mathbf{X}_i ; \boldsymbol{\mu}_{j'}, \boldsymbol{\Sigma}_{j'}\right)} . \label{eq:gamma_ij}
\end{align}
As the name suggests, the right-hand sides of the `updating' equations \eqref{eq:alpha_updating}, \eqref{eq:mu_updating}, \eqref{eq:sigma_updating} involve $\boldsymbol{\theta} = \left(\alpha_1,\ldots,\alpha_k, \boldsymbol{\mu}_1, \ldots, \boldsymbol{\mu}_k, \boldsymbol{\Sigma}_1,\ldots, \boldsymbol{\Sigma}_k \right)$ either explicitly or implicitly through $\gamma_{ij}$. As such, the updating equations are interlocking with each other and cannot be solved analytically.  Thus, by starting with an initial value for $\boldsymbol{\theta}$ on the right-hand sides of the updating equations, the left-hand sides are computed and plugged back to the right-hand sides iteratively until the convergence is reached. 
This optimization procedure is a version of the EM algorithm that alternates between the expectation step (computing $\gamma_{ij}$) and the maximization step (updating $\boldsymbol{\theta}$).  
}

A common challenge in using GMM to approximate the IS density is balancing between the estimated model's KL divergence (a closeness measure between the approximated IS density and the optimal one) and the model complexity. The studies \cite{botev2013, Wang2015, kurtz2013} that consider mixture models point out the difficulty associated with the choice of the number of mixture components, $k$. They either assume that $k$ is given \cite{botev2013, kurtz2013} or follow a rule of thumb based on ``some understanding of the structure of the problem at hand'' \cite{Wang2015}. {\color{black}More recently, an effective, but heuristic, clustering algorithm is also used to determine $k$ \cite{GEYER201915}}. Overcoming the need of picking $k$ a priori when using GMM to approximate the optimal DIS density, Choe \citep{Choe2017} proposes the cross-entropy information criterion (CIC) to select $k$ automatically \textcolor{black}{with a theoretical guarantee (\textit{i.e.}, CIC is an asymptotically unbiased estimator of the true cross-entropy).}


\vspace{-.3cm}

\subsection{Cross-Entropy Information Criterion}\label{sec:CIC}

In order to balance between the cross-entropy estimate and the model complexity, the CE estimator is minimized and the model complexity $k$ is concurrently penalized. CIC for deterministic simulation models takes the following form:
\begin{align}
\textrm{CIC}^{(t)}_{DIS}(d) &= \bar{\mathcal{C}}^{(t-1)}_{DIS}({\boldsymbol{\hat{\theta}}}) + \hat{K}^{(t-1)}_{_{DIS}}\frac{d}{\sum_{s=0}^{t-1} m^{(s)}} , \label{eq:CIC}
\end{align}
where {\color{black}$m^{(s)}$ is the number of input $\mathbf{X}$ sampled in iteration $s$ such that we have $\mathbf{X}_i^{(s)}, i=1,\ldots,m^{(s)}$.} The CE estimator $\bar{\mathcal{C}}^{(t-1)}_{DIS}({\boldsymbol{\hat{\theta}}})$ and $\hat{K}^{(t-1)}_{{DIS}}$ are calculated using all the aggregated data up to iteration $(t-1)$: {\color{black}
\begin{align}
\;\;\;\;\;\;\;\; \bar{\mathcal{C}}^{(t-1)}_{DIS}({\boldsymbol{\hat{\theta}}}) &=  - \frac{1}{\sum_{s=0}^{t-1} m^{(s)}}\sum_{s=0}^{t-1}\sum_{i=1}^{m^{(s)}} h(\mathbf{X}_i^{(s)}) w(\mathbf{X}_i^{(s)};\boldsymbol{\hat{\theta}}^{(s)}) \log{q\!\left(\mathbf{X}_i^{(s)} ; \boldsymbol{\hat{\theta}}\right)}   \label{eq:CE_est_hat_theta}
\end{align}
\begin{align}
\hat{K}^{(t-1)}_{{DIS}} &=  \frac{1}{\sum_{s=0}^{t-1} m^{(s)}}\sum_{s=0}^{t-1}\sum_{i=1}^{m^{(s)}} h(\mathbf{X}_i^{(s)}) w(\mathbf{X}_i^{(s)};\boldsymbol{\hat{\theta}}^{(s)}) . \label{eq:expect_h_w}
\end{align}}

 Note that $d = (k-1) + k(p+p(p+1)/2)$ is the dimension of $\boldsymbol{\hat{\theta}}$ and proportional to $k$, where $p$ denotes the dimension of the input $\mathbf{X}$. The second term of the CIC in \eqref{eq:CIC} penalizes the model complexity by being linearly proportional to $d$. The CIC is an asymptotically unbiased estimator of the cross-entropy (up to a multiplicative constant) under one key assumption that there exists $\boldsymbol{\theta}^*$ such that $q^*(\mathbf{x}) = q\!\left(\mathbf{x}; \boldsymbol{\theta}^*\right)$, and several other regularity conditions (see \citep{Choe2017} for details) that are similarly required for the Akaike information criterion (AIC) \citep{akaike1974} to be an asymptotically unbiased estimator of the true log-likelihood. Since for stochastic simulation models, $h(\mathbf{x})$ is unknown, it has to be estimated through $\hat h(\mathbf{x})$. We will present the CIC formulation for SIS in the next section.

\section{Methodology}\label{sec:method}
This section proposes a model selection solution to the estimation of the optimal IS density for stochastic simulation models. We use the GMM to express candidates for the optimal density. The CIC will automatically determine the mixture order $k$ of the GMM and control the model complexity. Among different mixture models, we choose the GMM since it is the most widely used mixture model thanks to its ability to model any smooth distribution. In general, the proposed CE method is not limited to just the GMM. It can be extended to other mixture models as long as we can minimize a CE estimator. In the convenient case of the GMM, an EM algorithm can be applied to estimate the GMM parameters. Moreover, it is computationally efficient to sample from the GMM.

It is natural to apply the concept of CIC in DIS to SIS. The expression of CIC involves $h(\mathbf{x})$, which is known in DIS but not in SIS. Hence we use the estimator $\hat h(\mathbf{x})$  which is a function of $\hat s(\mathbf{x})$, the estimated probability of failure at a particular $\mathbf{x}$ value.

\vspace{-.3cm}

\subsection{Approximations Necessary for Implementation}\label{sec:approx_implement}
{\color{black}

To implement the proposed method, we need to {\color{black} approximate $h(\mathbf{X}_i), i = 1, \ldots, m$,} 
for computing CIC, evaluating the EM algorithm equations (\ref{eq:alpha_updating})-(\ref{eq:gamma_ij}) to solve for $\boldsymbol{\hat{\theta}}$, and minimizing the CE estimator. For SIS, $h(\mathbf{x}) = \sqrt{	 s\!\left(\mathbf{x}\right)\left( 1- s\!\left(\mathbf{x}\right)\right)/n+ s\!\left(\mathbf{x}\right)^2 }$ \textcolor{black}{is unknown} 
because $s(\mathbf{x})$ is unknown. \textcolor{black}{Thus,} we estimate $s(\mathbf{X}_i)$ by
\begin{align}
\hat{s}(\mathbf{X}_i)  &= \frac{1}{N_i} \sum_{j=1}^{N_i} \mathbb{I}\left(Y_{j}^{(i)} > l\right) \label{eq:hat_s_plugin_est}
\end{align}
and then estimate {\color{black}$h(\mathbf{X}_i)$}  
by plugging in {\color{black}$\hat{s}(\mathbf{X}_i)$:}  
{\color{black}\begin{align}
\hat h(\mathbf{X}_i) = \sqrt{\hat s\!\left(\mathbf{X}_i\right)\left( 1- \hat s\!\left(\mathbf{X}_i\right)\right)/n + \hat s\!\left(\mathbf{X}_i\right)^2 }. \label{eq:hat_h}
\end{align}
We note that $h(\cdot)$ needs to be estimated only at the sampled $\mathbf{X}_i, i = 1, \ldots, m$, for which we already have run the simulator $N_i$ times and have the evaluation of $\hat{s}(\mathbf{X}_i)$  in \eqref{eq:hat_s_plugin_est} to ultimately compute the estimator in \eqref{eq:SIS_estimator} for the failure probability. Thus, by utilizing the already available information, this approximation of $h(\mathbf{X}_i), i = 1, \ldots, m$, does not incur an extra computational cost.}

{\color{black}  
As the sample size $n$ tends to infinity while $m$ is allowed to grow slowly than $n$ (or as a special case, $m$ may be fixed), $\hat h(\mathbf{X}_i)$ converges to $h(\mathbf{X}_i)$ in probability for any $i$. Specifically, because $N_i = O_p(n/m)$, if $n$ increases faster than $m$ (\textit{i.e.}, $m/n=o(1)$),  $\hat{s}(\mathbf{X}_i)$ in \eqref{eq:hat_s_plugin_est} converges in probability to $s(\mathbf{X}_i)$ for any $i$ by the law of large numbers. Therefore, $\hat h(\mathbf{X}_i)$ in \eqref{eq:hat_h} converges to $h(\mathbf{X}_i)$ in probability for any $i$ as $n\to\infty$, by the continuous mapping theorem.}

SIS also needs to allocate $N_i$ replications at each $\mathbf{X}_i$, as explained in Section~\ref{sec:background}. In Appendix 1, we show that for a large $n \gg \max_{i=1}^{m} {(1-s\!\left(\mathbf{X}_i\right))}/{s\!\left(\mathbf{X}_i\right)}$,  the optimal $N_i$ in \eqref{eq:N_i_opt} is approximately proportional to $\sqrt{w(\mathbf{X}_i) - \hat{P}_{SIS}}$. Thus, we decide $N_i$ based on this approximation. If $w(\mathbf{X}_i) - \hat{P}_{SIS} \leq 0$, we assign $N_i = 1$, to ensure the unbiasedness of $\hat{P}_{SIS}$ in \eqref{eq:SIS_estimator}. 
{\color{black}We note that such approximation of $N_i$ would not substantially affect the performance of SIS because our numerical study results in Section~\ref{sec:num_ex} and other SIS implementation results in \cite{Choe2015} suggest that SIS is rather insensitive to the ratio of $m$ to $n$, which determines the magnitude of order of $N_i$.} 
}

\vspace{-.3cm}

\subsection{Aggregated Failure Probability Estimation}

We aggregately use the samples obtained in all of the CE iterations. Instead of $\hat{P}_{SIS}$ in \eqref{eq:SIS_estimator}, we compute the failure probability estimator $\bar{P}_{SIS}^{(t)}$ using all the aggregated data up to iteration $t$: 
\begin{align}
\bar{P}_{SIS}^{(t)} &= \frac{1}{t+1} \sum_{s = 0}^{t} \frac{1}{m^{(s)}} \sum_{i=1}^{m^{(s)}} \frac{1}{N_i^{(s)}} \sum_{j=1}^{N_i^{(s)}} \mathbb{I}\!\left(Y_{ij}^{(s)} > l\right) \frac{f\!\left(\mathbf{X}_{i}^{(s)}\right)}{q\!\left(\mathbf{X}_{i}^{(s)}; \boldsymbol{\hat{\theta}}^{(s)}\right)},
\label{eq:agg_SIS_estimator}
\end{align}
where $\mathbf{X}_{i}^{(s)}$ is the $i^{th}$ random input at iteration $s$, $m^{(s)}$ is the number of $\mathbf{X}_{i}^{(s)}$ drawn at iteration $s$, and ${N_i^{(s)}}$ is the number of simulation runs at each $\mathbf{X}_{i}^{(s)}$.


The idea of data aggregation leads to the aggregated CIC formulation for stochastic simulation models as follows: 
\begin{align}
\textrm{CIC}^{(t)}_{SIS}(d) &= \bar{\mathcal{C}}^{(t-1)}_{SIS}({\boldsymbol{\hat{\theta}}}) + \hat{K}^{(t-1)}_{{SIS}}\frac{d}{\sum_{s=0}^{t-1} m^{(s)}} , \label{eq:CIC_SIS}
\end{align}
where the CE estimator $\bar{\mathcal{C}}^{(t-1)}_{SIS}({\boldsymbol{\hat{\theta}}})$ and $\hat{K}^{(t-1)}_{{SIS}}$ are calculated using all the aggregated data up to iteration $(t-1)$:
{\color{black}
\begin{align}
\;\;\;\;\;\;\;\; \bar{\mathcal{C}}^{(t-1)}_{SIS}({\boldsymbol{\hat{\theta}}}) &=  - \frac{1}{\sum_{s=0}^{t-1} m^{(s)}}\sum_{s=0}^{t-1}\sum_{i=1}^{m^{(s)}} \hat h(\mathbf{X}_i^{(s)}) w(\mathbf{X}_i^{(s)}; \boldsymbol{\hat{\theta}}^{(s)}) \log{q\!\left(\mathbf{X}_i^{(s)} ; \boldsymbol{\hat{\theta}}\right)}  . \label{eq:CE_SIS_est_hat_theta}
\end{align}}

Noting that $\hat K^{(t-1)}_{DIS}$ in \eqref{eq:CIC} is an unbiased DIS estimator of the failure probability, we similarly set 
\begin{align}
\hat{K}^{(t-1)}_{{SIS}} &= \bar{P}_{SIS}^{(t-1)}.  \label{eq:expect_h_w_SIS}
\end{align}

\vspace{-.3cm}

\subsection{Summary of the Proposed Method (CE-SIS) \textcolor{black}{and Implementation Guideline}}\label{eq:EMCE_algo}

For the EM initialization, the grid search of the optimal number of mixture components $k^{(t)*}$, the stopping criteria of the EM algorithm to minimize the quantity $\bar{\mathcal{C}}^{(t-1)}_{SIS}({\boldsymbol{\hat{\theta}}})$ in \eqref{eq:CE_SIS_est_hat_theta}, and whole division of the replication allocation $N_i^{(t)}$, we follow the same procedures in \citep{Choe2017}. The details of implementation are provided in Appendix 2. 


We propose the CE-SIS pseudo-code in Algorithm~\ref{alg:alg1}: 

\textbf{INPUT:}
\begin{itemize}
	\item Initialize the GMM $q(\mathbf{x}; \boldsymbol{\hat{\theta}}^{(0)})$ for the first iteration (e.g., GMM with $k = 1$ and the identity covariance matrix).
	\item Given a simulation budget, $n$, we pick the actual simulation runs at each iteration, $n^{(t)}$, such that $\sum_{t=0}^{\tau} n ^{(t)} = n$, where $\tau$ denotes the \textcolor{black}{total number of CE iterations}.
	\item Fix the number of $\mathbf{X}$ values to draw, $m^{(t)}$, for each iteration $t \geq 1$ such that, for example, $m^{(t)} \approx 0.3n^{(t)}$, as adopted in \citep{Choe2015}. According to \citep{Choe2015}, SIS should be generally insensitive to the ratio $\frac{m^{(t)}}{n^{(t)}}$. Note that at the first iteration $t=0$, we set $m^{(0)} = n^{(0)}$ or equivalently $N_i^{(0)} = 1$, to explore the input space as much as possible. 
\end{itemize}

\begin{algorithm}[!htbp]

\caption{Approximating the optimal IS density for SIS}
\begin{algorithmic} 
\label{alg:alg1}

\STATE $t = 0$
\WHILE{$t \leq \tau $}
	\IF{$t = 0$}
    	\STATE $N_i^{(0)} = 1$ for $i \in \{1,  \ldots, m^{(0)}\}$
		\FOR{$i \in \{1,  \ldots, m^{(0)}\}$}       	
			\STATE {Sample $\mathbf{X}_{i}^{(0)}$ from $q(\mathbf{x}; \boldsymbol{\hat{\theta}}^{(0)})$}, an initial GMM
    		\STATE {Generate the data using the simulator: $\mathcal{D} = \mathcal{D}^{(0)} = \{(\mathbf{X}_{i}^{(0)}, Y_{i1}^{(0)})\}$}
		\ENDFOR\\
		Compute $\bar{P}_{SIS}^{(0)}$ in \eqref{eq:agg_SIS_estimator}
	\ELSE 
		\FOR{$k \in \{k_{\min}^{(t)},  \ldots, k_{\max}^{(t)}\}$}
			\STATE {Using EM: $\boldsymbol{\hat \theta}^{(t)}\!(k) = \arg\!\min_{\boldsymbol{\hat \theta}(k) \in \boldsymbol{\Theta}(d(k))}\bar{\mathcal{C}}^{(t-1)}_{SIS}({\boldsymbol{\hat{\theta}}}(k))$ in \eqref{eq:CE_SIS_est_hat_theta}}
			\STATE {Let $\hat{K}^{(t-1)}_{{SIS}}\!(k) = \bar{P}_{SIS}^{(t-1)}\!(k)$} in \eqref{eq:expect_h_w_SIS}
			\STATE {Compute $\textrm{CIC}^{(t)}_{SIS}(d(k))$ in \eqref{eq:CIC_SIS}}
		\ENDFOR\\
		Pick $k^{(t)*} = \arg\!\min_{k}\textrm{CIC}^{(t)}_{SIS}(d(k))$\\
		\STATE $\boldsymbol{\hat \theta}^{(t)} \gets \boldsymbol{\hat \theta}{(k^{(t)*})}$

		\FOR{$i \in \{1,  \ldots, m^{(t)}\}$}
			\STATE {Sample $\mathbf{X}_{i}^{(t)}$ from $q(\mathbf{x}; \boldsymbol{\hat{\theta}}^{(t)})$}
			\STATE {Compute ${\sqrt{\left[w(\mathbf{X}^{(t)}_i) - \bar{P}_{SIS}^{(t-1)}\right]_+}}$}		
		\ENDFOR
		\FOR{$i \in \{1,  \ldots, m^{(t)}\}$}
			\STATE {Compute $N_i^{(t)} = max\left( 1, (round){n^{(t)}\frac{\sqrt{\left[w(\mathbf{X}^{(t)}_i) - \bar{P}_{SIS}^{(t-1)}\right]_+}}{\sum_{i=1}^{m^{(t)}}{\sqrt{\left[w(\mathbf{X}^{(t)}_i) - \bar{P}_{SIS}^{(t-1)}\right]_+}}}} \right) $}
		\ENDFOR
        \FOR{$i \in \{1,  \ldots, m^{(t)}\}$}
			\FOR{$j\in\{1,\ldots, N_i^{(t)}\}$} 
    			\STATE {Generate the data using the simulator: $\mathcal{D}^{(t)} = \{(\mathbf{X}_{i}^{(t)}, Y_{ij}^{(t)})\}$}
   			\ENDFOR
        \ENDFOR
		\STATE {Aggregate data: $\mathcal{D} = \bigcup_{s=0}^{t} \mathcal{D}^{(s)}$}\\
		Compute $\bar{P}_{SIS}^{(t)}$ in \eqref{eq:agg_SIS_estimator} \\		
	\ENDIF	
	\STATE $t\gets t+1$
\ENDWHILE
\end{algorithmic}
\end{algorithm}

\begin{figure}[!h]
{\centering
{\color{black}\subfigure[Initial pilot sample: Uniform $(-5,5)$.]{\includegraphics[width = 5.5cm]{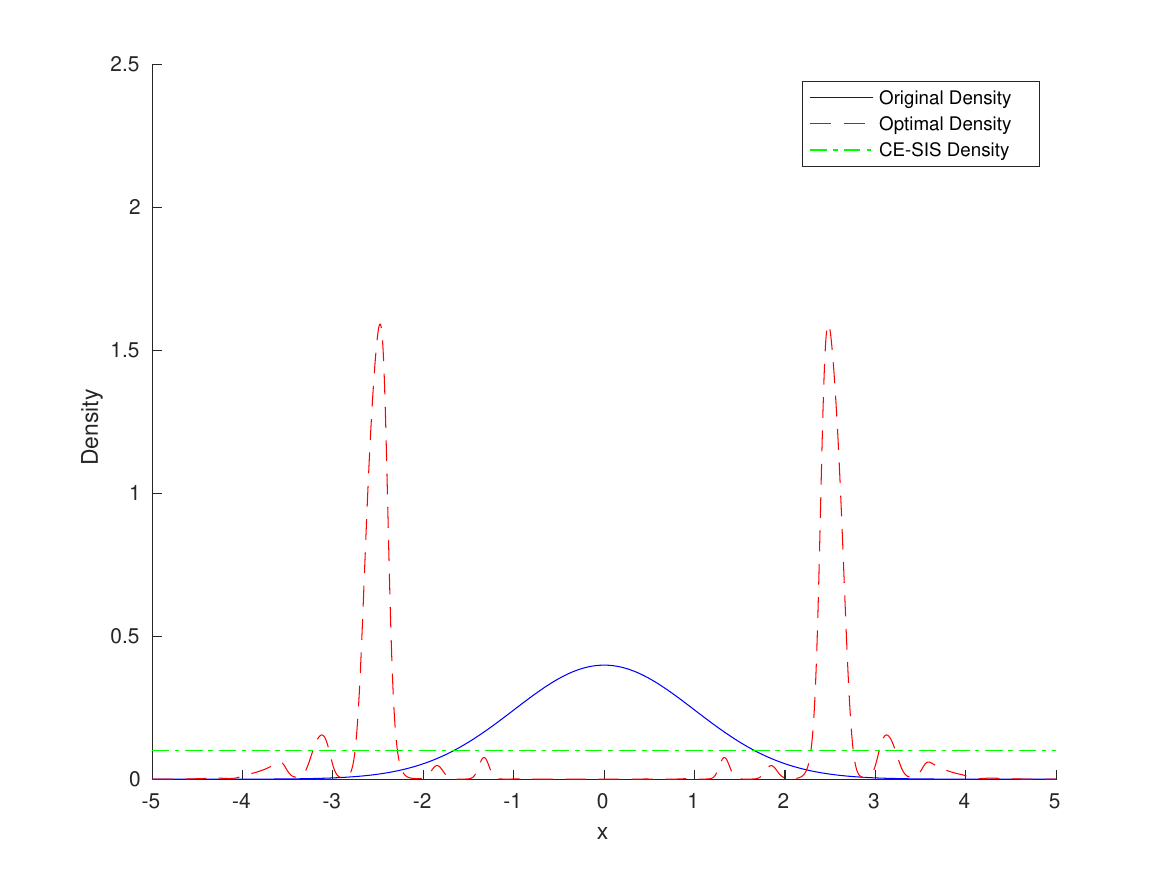}}
\subfigure[1st iteration: 4 GMM components.]{\includegraphics[width = 5.5cm]{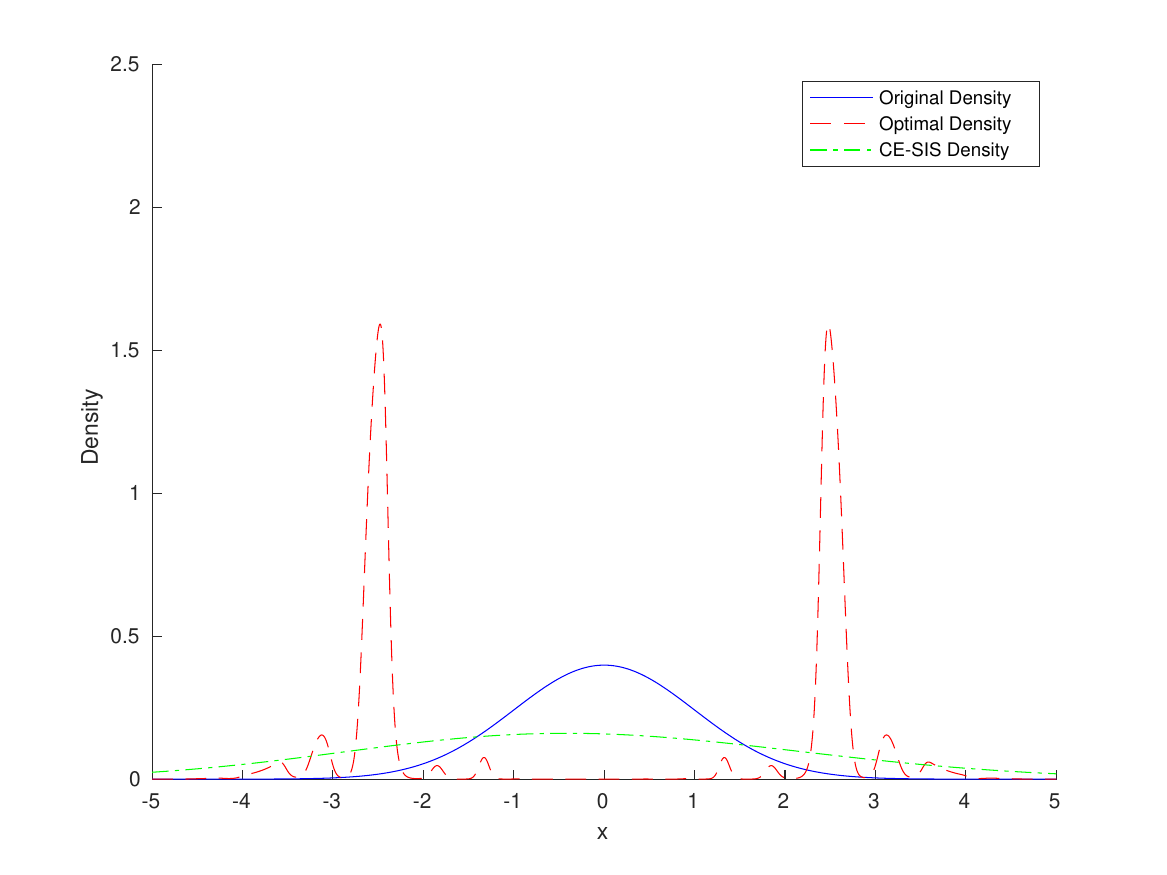}}
\subfigure[3rd iteration: 7 GMM components.]{\includegraphics[width = 5.5cm]{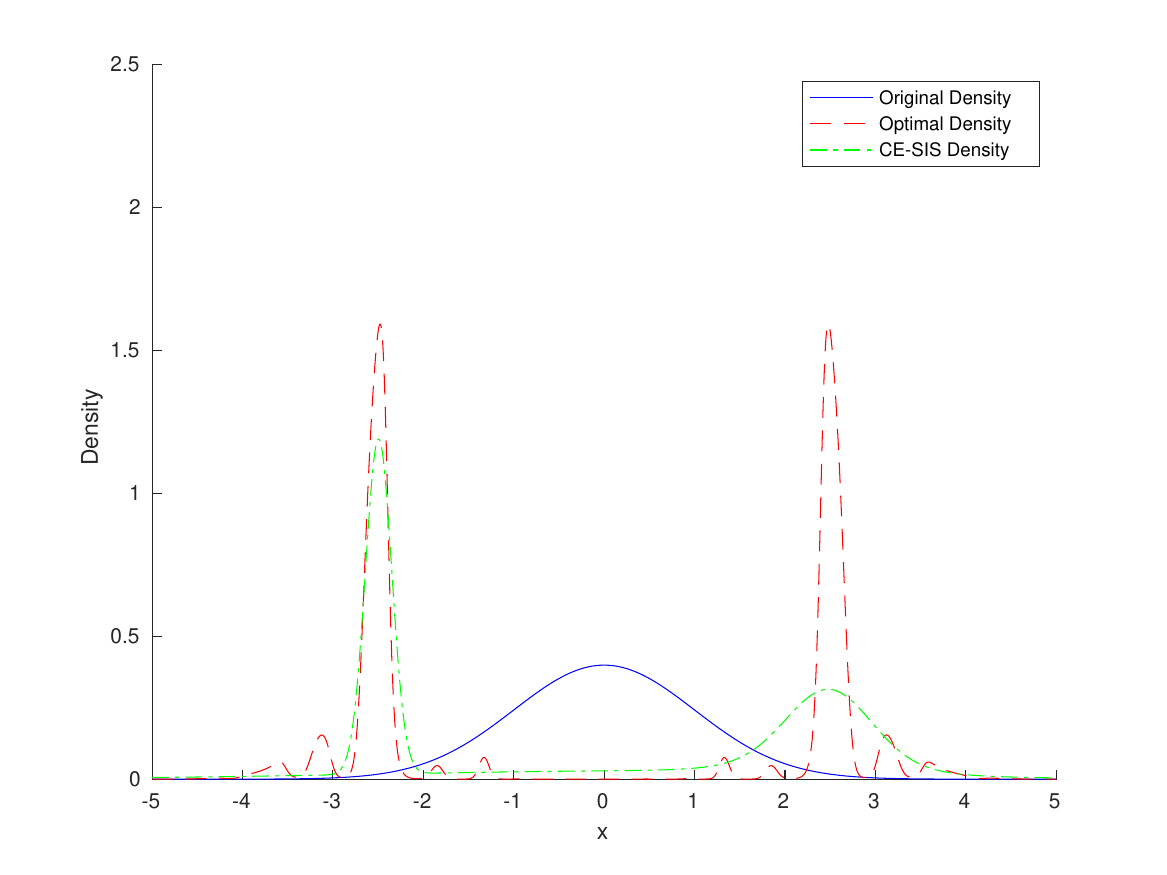}}
\subfigure[6th iteration: 6 GMM components.]{\includegraphics[width = 5.5cm]{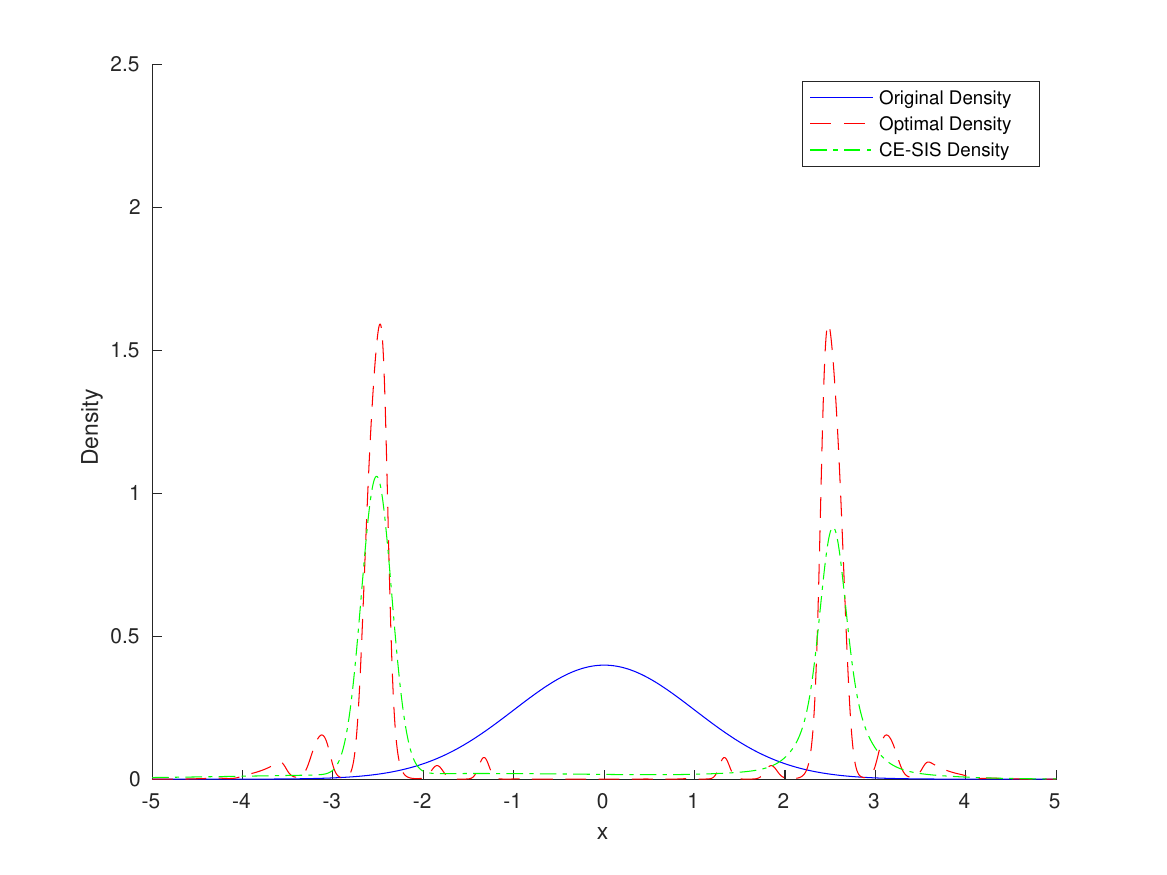}}
\subfigure[9th iteration: 8 GMM components.]{\includegraphics[width = 5.5cm]{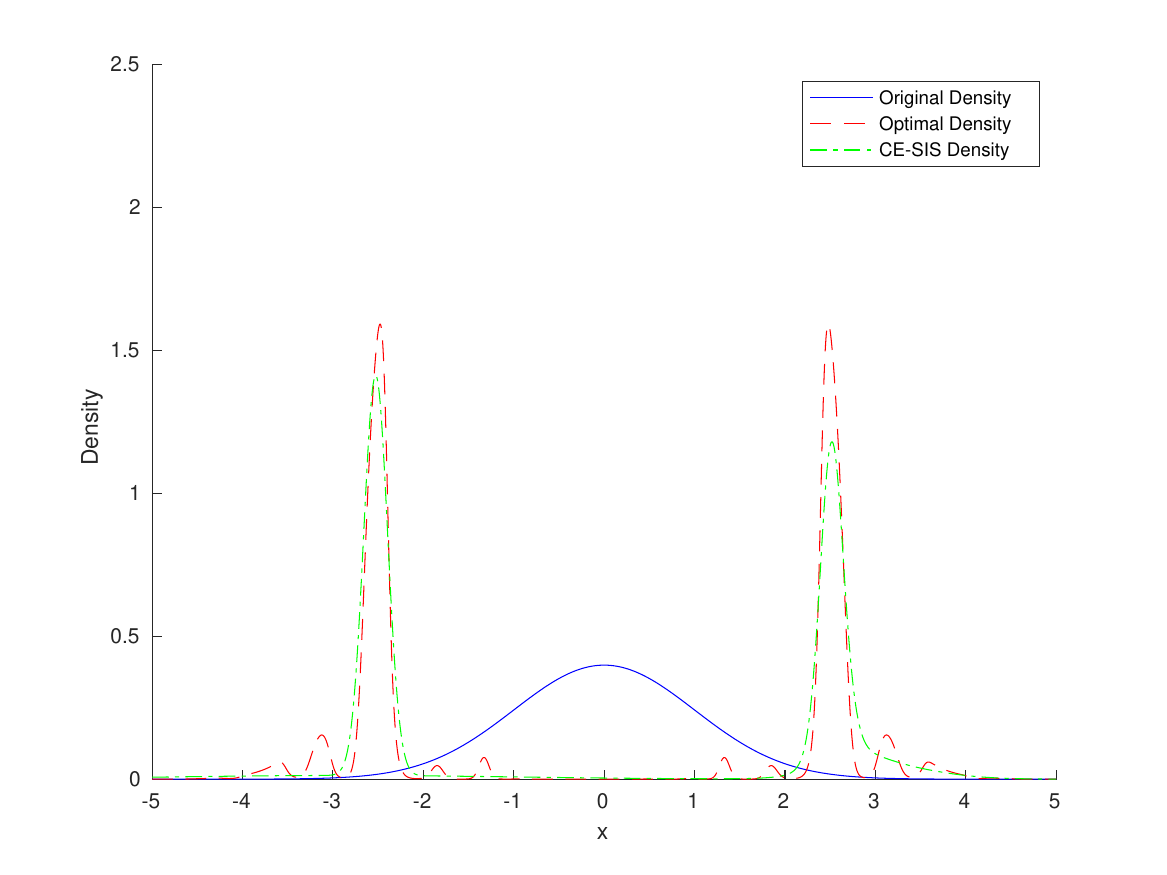}}
\subfigure[10th (\& final) iteration: 9 GMM components.]{\includegraphics[width = 5.5cm]{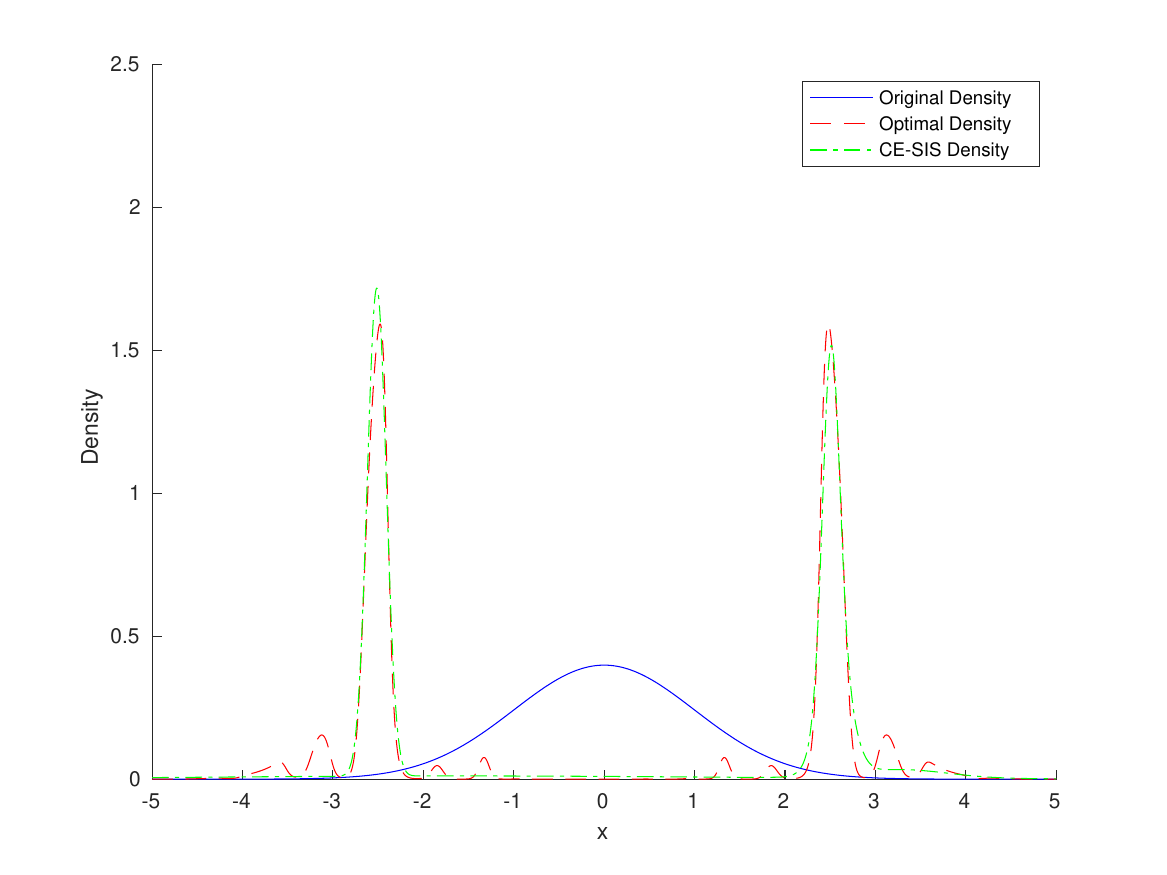}}}

\caption{{\color{black}Evolution of the CE-SIS density (in green dash-dotted line) that converges to the optimal SIS density $q_{SIS}$ in \eqref{eq:SIS_opt_density} (in red dashed line) within one simulation experiment for Cannemela example in Section~\ref{sec:num_ex} where the failure probability is 0.01 and the original input density $f$ (in black solid line) is the standard normal density.}}
\label{fig:evolution}
}
\end{figure}

{\color{black}Figure~\ref{fig:evolution} shows the evolution of the sampling density of $X$ over iterations $t=0,1,3,6,9,10$ in a numerical example in Section~\ref{sec:num_ex}. The CE-SIS density (in green dash-dotted line) converges to the optimal SIS density $q_{SIS}$ in \eqref{eq:SIS_opt_density} (in red dashed line) over iterations. At $t=0$ in the pilot run, $X$ is sampled from a uniform distribution $(-5,5)$, instead of a GMM, to reflect no knowledge of important region in the example. After about 6 iterations, the GMM starts picking up the important region effectively.}

{\color{black} Algorithm~\ref{alg:alg1} assumes that the user of the algorithm can set an initial GMM $q(\mathbf{x}; \boldsymbol{\hat{\theta}}^{(0)})$ such that the mass of the density well covers potential failure regions. It ensures that the failure probability estimator $\bar{P}_{SIS}^{(0)}$ at iteration 0 is non-zero with a sufficiently large $m^{(0)}$. If the user has no knowledge of potential failure regions, the user can adapt the algorithm by adopting the \textit{multilevel} CE procedure widely used in the literature \cite{rubinstein2013cross,rubinstein2016simulation}. Specifically, the failure threshold level $l$ can be adjusted over iterations so that the failure probability estimator can remain non-zero for the series of thresholds which approach the target threshold $l$ over iterations. To keep Algorithm~\ref{alg:alg1} simple and convey the main idea of the proposed method, we do not present a multilevel-procedure algorithm, but we refer an interested reader to the following references \cite{rubinstein2013cross,rubinstein2016simulation}.}

{\color{black} The total number of CE iterations, $\tau$, is set based on the simulation budget $n = \sum_{t=0}^{\tau} n ^{(t)}$ that a user can afford. However, the user can exercise the freedom to stop the algorithm early, for example, if the estimated variance of the failure probability estimator in \eqref{eq:agg_SIS_estimator} is small enough compared to the probability estimate (\textit{e.g.}, the coefficient of variation is 20\% or less). The user can use a consistent estimator of the variance derived in \cite{choe2018uncertainty}. On the other hand, if the user realizes that a larger $\tau$ is needed to reduce the coefficient of variation, then the algorithm can be set to continue to the next iteration.}

{\color{black} At any iteration $t$ in Algorithm~\ref{alg:alg1}, $k_{\min}^{(t)}$ can be always set to be 1 and $k_{\max}^{(t)}$ can be always set to be the maximum number that yields the maximum dimension $d$ of the parameter $\boldsymbol{{\theta}}$. Specifically, recall that $d$ is equal to $(k-1) + k(p+p(p+1)/2)$ for a GMM with $k$ components, where $p$ denotes the dimension of the input $\mathbf{X}$. To estimate $\boldsymbol{{\theta}}$, the total sample size $n$ should be at least $d$, that is, $n\ge d$. Because $k = (d+1) / (p+p(p+1)/2 +1)$,  we can set $k_{\max}^{(t)}$ to be  $(n+1)/(p+p(p+1)/2 +1)$. In practice, the grid search over $k \in \{k_{\min}^{(t)},  \ldots, k_{\max}^{(t)}\}$ to minimize $\textrm{CIC}^{(t)}_{SIS}(d(k))$ does not require increasing $k$ up to $k_{\max}^{(t)}$. For algorithmic efficiency, we take a heuristic approach from \cite{Choe2017} that uses the moving average (with the window size of four) of the CIC to stop increasing $k$ when the moving average starts to increase.}

{\color{black} In practice, the major computational cost of Algorithm~\ref{alg:alg1} lies in running the (high-fidelity) simulator to obtain $Y_{ij}^{(t)}$. Compared to this cost (e.g., roughly 1 minute per run in our case study), the running times of iterative procedures in the algorithm are negligible. For example, running the EM algorithm, minimizing the CIC, and computing $N_i^{(t)}, 1,  \ldots, m^{(t)}$, in each iteration take seconds.}

\section{\textcolor{black}{Numerical Studies}}\label{sec:num_ex}

{\color{black}In this section, we conduct extensive numerical studies to investigate the performances and sensitivities of the CE-SIS algorithm with respect to five aspects: target failure probability to estimate (Table~\ref{tab:ex2-probability}), $\frac{m}{n}$ ratio (Table~\ref{tab:ex2-ratio}), distribution of univariate input $X$ (Table~\ref{tab:ex2-distribution}), dimension of multivariate input $\mathbf{X}$ (Table~\ref{tab:ex2-high}), and standard deviation of $Y|\mathbf{X}$ (Table~\ref{tab:ex2-high}).
We use two numerical examples from the literature \cite{choe2018uncertainty,chen2017oracle} for which metamodel-based SIS methods are already applied. Assuming that the data-generating processes are unknown, we compare the performance of the CE-SIS algorithm with the metamodel-based approaches. Alongside we also present the results from the optimal SIS algorithm (using the optimal density in \eqref{eq:SIS_opt_density} and the allocation in \eqref{eq:N_i_opt}) that utilizes the full knowledge of the data-generating processes (\textit{e.g.}, the conditional probability $s(X) = \mathbb{P}(Y > l \mid X)$).}


{\color{black}\subsection{Cannamela Example: Univariate Input}}
{\color{black}As a numerical example with the univariate input $X$, we use an example from \cite{Choe2015} that modified a deterministic simulation model example originally from Cannamela et al. \cite{cannamela2008} into a stochastic simulation model example}. 
Its data generating structure is as follows:
\begin{align}
X \sim \mathcal{N}\!\left(0,1\right), \;\;\;\;\;\;
& Y \vert X \sim \mathcal{N}\!\left(\mu\!\left(X\right),\sigma^2\!\left(X\right)\right),  \nonumber
\end{align}
where the mean and the standard deviation are:
\begin{align}
\mu\!\left(X\right) &=0.95  X^2 \left(1+0.5 \cos(5 X)+0.5 \cos(10  X) \right), \nonumber
\\
\sigma\!\left(X\right) &= 1+0.7\left|X\right| + 0.4 \cos(X)+0.3\cos(14X).  \nonumber
\end{align}

{\color{black}The metamodel-based SIS method in \cite{choe2018uncertainty} constructs a metamodel of $Y|X$ and uses it to estimate the conditional probability $s(X) = \mathbb{P}(Y > l \mid X)$, which is in turn needed to approximate the optimal density in \eqref{eq:SIS_opt_density} and the allocation in \eqref{eq:N_i_opt}. This metamodel \textit{assumes} the knowledge that $Y|X$ follows a normal distribution. This assumption is strong since such distribution is typically \textit{unknown} for a stochastic simulation model in practice. With the distributional assumption, the metamodel 
approximates the parameters, mean $\mu\!\left(X\right)$ and standard deviation $\sigma\!\left(X\right)$ of $Y|X$, using cubic spline functions of $X$ in the generalized additive model for location, scale and shape (GAMLSS) framework \cite{rigby2005}. An interested reader is referred to \cite{choe2018uncertainty} for the modeling details.  The metamodel is built using a pilot dataset of 3000 simulation replications and the IS experiment is based on 10,000 simulation replications. To ensure that both methods use the same simulation budget, for the CE-SIS method, we use $n^{(0)} = 3000$ and $n^{(t)} = 1000$ for $t=1,\ldots,10$. For the pilot/initial data of both methods, we sample $X$ from the uniform distribution $(-5,5)$ assuming no knowledge of important regions (akin to an uninformative prior in Bayesian inference). }

A performance metric we use is the CMC ratio, which is defined as
\[
\textrm{CMC ratio} = \frac{n} {n_{CMC}},
\]
where $n$ is the total number of simulation replications (e.g., 3000 + 10000 for metamodel-based SIS, 3000 + 10000 for CE-SIS, and 10000 for the optimal SIS since we do not need to build the metamodel for the optimal SIS.) and $n_{CMC}$ is the total number of simulation replications required for the CMC simulation to achieve the same standard error as each method, which is calculated as
\[
n_{CMC} = \frac{{P}(1-{P})}{S.E.^2}.
\]
{\color{black}Note that $S.E.$ is the standard error of each method. ${P}$ is the target failure probability $\mathbb{P}\left( Y >l \right)$ we would like to estimate. The smaller CMC ratio translates to a smaller number of replications used in each row's method divided by the number of replications necessary for CMC in \eqref{eq:CMC_estimator} to achieve the standard error in the row, which means better computational saving for the same accuracy level.}

{\color{black} We determine the threshold $l$ for the target failure probability $\mathbb{P}\left( Y >l \right)$ using a large-scale Monte Carlo simulation (with $10^7$ replications) so that $l$ yields the target failure probability.}

{\color{black}The simulation experiment is repeated $500$ times to obtain the results. Using this example, we study performances of the CE-SIS algorithm with respect to the following three aspects:
\begin{itemize}
    \item Target failure probability $\mathbb{P}\left( Y >l \right)$: $0.01, 0.001, 0.0001$ in Table~\ref{tab:ex2-probability}.
    \item $\frac{m}{n}$ ratio: 0.1, 0.3, 0.5, 0.7, 0.9 in Table~\ref{tab:ex2-ratio}. Recall that $m$ is the number of distinct $X$ values and $n$ is the total number of simulation runs. 
    \item Distribution of $X$: 
    \begin{itemize}
        \item Normal distribution with mean 0 and standard deviation 1,
        \item $t$-distribution with the number of degrees of freedom ($df$) equal to 5 or 10,
        \item Weibull distribution with scale parameter of 1 and shape parameter of 1 or 1.5
    \end{itemize} 
    in Table~\ref{tab:ex2-distribution}. Both non-normal distributions have heavier tails than the normal distribution, especially heavier when $df$ is 5 for the $t$-distribution and the shape parameter is 1 for the Weibull distribution. 
\end{itemize}}

{\color{black}
As shown in Table~\ref{tab:ex2-probability}, CE-SIS provides significant computational saving over CMC. The saving is larger when we estimate the probability of a rarer event, as expected of an IS scheme \cite{Choe2015}.  However, the CE-SIS underperforms the metamodel-based approach, which assumes the knowledge of the conditional distribution of $Y|X$.  We remark that  the performance of metamodel-based SIS can be made arbitrarily better or worse by changing the metamodel. For instance, if the metamodel of $s(X)$ is a constant over $X$ (\textit{i.e.}, no knowledge of which region is more important), then the metamodel-based SIS density (see \eqref{eq:SIS_opt_density}) reduces to the original input density $f$ so that the performance of SIS becomes equal to that of CMC. Thus, we instead use a good metamodel from the literature and present the result as a point of reference. In reality, the underlying data-generating distribution is unknown and likely much more complex, rendering the construction of a good metamodel difficult, especially when the pilot sample size is limited. In contrast, the CE-SIS method can automatically and conveniently yield a good SIS density. 

From Table~\ref{tab:ex2-ratio} and Table~\ref{tab:ex2-distribution}, it is also observed that the CE-SIS method is generally robust to different $\frac{m}{n}$ ratios and distributions of $X$, in comparison to the metamodel-based SIS. The marked exceptions occur when the original input density $f$ is highly heavy-tailed: $t(df=5)$ and $Weibull(1,1)$. 
Such large standard errors and CMC ratios are expected because the optimal SIS density in \eqref{eq:SIS_opt_density} is proportional to $f$ and thus heavy-tailed for the given $f$. Any GMM (\textit{i.e.}, a finite mixture of Gaussian distributions) cannot model heavy-tailed distributions by definition of heavy-tailedness. A good news is that in practice, users of the CE-SIS algorithm always know whether $f$ is heavy-tailed or not. Thus, they can work around the issue (\textit{e.g.}, use a mixture of heavy-tailed distributions such as $t$ distributions).}


\begin{table}
\centering
{\color{black}\caption{Comparison between CE-SIS, metamodel-based SIS, and optimal SIS with respect to different target probability $\mathbb{P}\left( Y >l \right)$ to estimate. $\frac{m}{n}$ ratio is fixed at 0.3. The results are based on 500 experiments.}}
\label{tab:ex2-probability}
{\color{black}\begin{tabular}{cccccc }
\hline
$\mathbb{P}\left( Y >l \right)$& Threshold $l$ &Method     & Mean    & S.E. & CMC Ratio \\ \hline\hline
0.010000                 & 9.13   & CE-SIS       & 0.009984&	0.000391        & 20.08\%      \\
                                &   & Metamodel  & 0.010007	&0.000292    & 11.20\%     \\ 
                                & & Optimal SIS  & 0.009994	&0.000229     & 5.30\%     \\ 
0.001000                 & 14.60   & CE-SIS       & 0.001001&	0.000097    & 12.27\%      \\
                &                   & Metamodel  & 0.001001&	0.000031       & 1.22\%     \\
                                    & & Optimal SIS  & 0.001000	&0.000023     & 0.51\%     \\
0.000100                 & 24.29   & CE-SIS       & 0.000100&	0.000005        &0.35\%      \\
                                &   & Metamodel  & 0.000100&0.000001       & 0.01\%     \\
                                    & & Optimal SIS  & 0.000102&	0.000002     & 0.02\%     \\
\hline
\end{tabular}}
\end{table}

\begin{table}
\centering
{\color{black}\caption{Comparison between CE-SIS, metamodel-based SIS, and optimal SIS with respect to different $\frac{m}{n}$ ratios. The target probability $\mathbb{P}\left( Y >l \right)$ to estimate is fixed at 0.01. The results are based on 500 experiments.}}
\label{tab:ex2-ratio}
{\color{black}\begin{tabular}{cccccc}
\hline
$\frac{m}{n}$ ratio & Threshold $l$ &Method     & Mean    & S.E. & CMC Ratio \\ \hline\hline
0.1                 & 9.13   & CE-SIS        & 0.009975&	0.000371       & 18.09\%      \\
                         &   & Metamodel     & 0.010013&	0.000338        & 15.02\%      \\
                         &   & Optimal SIS     & 0.010015&	0.000234       & 5.52\%      \\                 
0.3                 & 9.13   & CE-SIS         & 0.009984&	0.000391     & 20.08\%     \\ 
                         &    & Metamodel     & 0.010007&	0.000292        & 11.20\%      \\
                         &   & Optimal SIS     &0.009994&	0.000229       & 5.30\%      \\   
0.5                 & 9.13   & CE-SIS        & 0.009979&	0.000493     & 31.93\%      \\
                            &   & Metamodel     & 0.010014&	0.000213        & 5.93\%      \\
                         &   & Optimal SIS     & 0.009980&	0.000233        & 5.48\%      \\   
0.7                 & 9.13   & CE-SIS         & 0.009993&	0.000357        & 16.70\%     \\
                            &   & Metamodel     & 0.010009&	0.000215        & 6.07\%      \\
                         &   & Optimal SIS     & 0.009992&	0.000237       & 5.65\%      \\   
0.9                 & 9.13   & CE-SIS        & 0.009966&	0.000495       &32.13\%      \\
                            &    & Metamodel     & 0.009998&0.000206       & 5.58\%      \\
                         &   & Optimal SIS     & 0.010022&	0.000231       & 5.37\%      \\   
\hline
\end{tabular}}
\end{table}

\begin{table}
\centering
{\color{black}\caption{Comparison between CE-SIS, metamodel-based SIS, and optimal SIS with respect to different distributions of $X$. The target probability $\mathbb{P}\left( Y >l \right)$ to estimate is fixed at 0.01. $\frac{m}{n}$ ratio is fixed at 0.3. The results are based on 500 experiments.}}
\label{tab:ex2-distribution}
{\color{black}\begin{tabular}{cccccc}
\hline
$X$ Distribution & Threshold $l$ &Method     & Mean    & S.E. & CMC Ratio \\ \hline\hline
$\mathcal{N}\!\left(0,1\right)$                & 9.13   & CE-SIS         & 0.009984&	0.000391       & 20.08\%      \\
                         &    & Metamodel      & 0.010007&	0.000292        & 11.20\%      \\
                         &    & Optimal SIS      & 0.009994&	0.000229        & 5.30\%      \\
$t(df=5)$                 & 16.97  & CE-SIS           & 0.009457&	0.001477    & 286.28\%     \\ 
                         &    & Metamodel       & 0.009991&	0.000076        & 0.77\%      \\
                         &    & Optimal SIS      & 0.009995&	0.000097       & 0.95\%      \\           
$t(df=10)$                 & 12.02  & CE-SIS          & 0.010002&	0.000377     & 18.64\%     \\ 
                            &    & Metamodel      & 0.010005&	0.000132        & 2.30\%      \\
                         &    & Optimal SIS      & 0.010002&	0.000121        & 1.48\%      \\
$Weibull(1,1)$                 & 23.57   & CE-SIS        & 0.008840&	0.001883     & 465.66\%      \\
                        &    & Metamodel     & 0.010002&	0.000108       & 1.52\%      \\
                         &    & Optimal SIS      & 0.009999&	0.000057        & 0.33\%      \\
$Weibull(1,1.5)$                & 10.38   & CE-SIS      & 0.009964&	0.000268     & 9.41\%      \\
                        &    & Metamodel   &0.009994&	0.000195       & 4.99\%      \\
                         &    & Optimal SIS      & 0.010000&	0.000146       & 2.15\%      \\

\hline
\end{tabular}}
\end{table}

{\color{black}\subsection{Ackley Example: Multivariate Input}}
{\color{black}As a numerical example with the multivariate input $\mathbf{X} = (X_{1}, \ldots, X_{d}) \in \mathbb{R}^d$, we use an example from \cite{chen2017oracle} that modified an example in Ackley \cite{ackley1987} as follows:
\begin{align}
\mathbf{X} \sim \mathcal{N}\!\left(\mathbf{0}, \mathbf{I}\right), \;\;\;\;\;\;
& Y \vert \mathbf{X} \sim \mathcal{N}\!\left(\mu\!\left(\mathbf{X}\right),1\right),  \nonumber
\end{align}
where the mean is:
\begin{align}
\mu\!\left(\mathbf{X}\right) &=20\left(1-\exp\left(-0.2\sqrt{\frac{1}{d}\left\Vert \mathbf{X}\right\Vert^2}\right)\right) + \left(\exp(1) - \exp\left(\frac{1}{d}\sum_{i=1}^{d}\cos(2\pi {X}_i)\right)\right). \nonumber
\end{align}}

{\color{black}For the construction of a metamodel, we adopt the same metamodel-based approach in \cite{chen2017oracle} that fits a logistic regression model on a pilot dataset to estimate $s\!\left(\mathbf{X}\right) = \mathbb{P}\!\left(Y >l \mid \mathbf{X} \right)$. However, their model $s_\theta(\mathbf{X}) = \left( 1 + e^{\theta_0 + \theta_1 X_1 + \ldots + \theta_d X_d} \right)^{-1}$  considers only linear terms of $\mathbf{X}=(X_1,\ldots, X_d)$ and cannot well model $s\!\left(\mathbf{X}\right)$ that has a symmetry with respect to the origin (see $\mu\!\left(\mathbf{X}\right)$). Accordingly, the resulting metamodel-based SIS tends to yield the standard error as large as CMC \cite{chen2017oracle}. Thus, assuming the knowledge of the symmetric structure, we add quadratic terms of $\mathbf{X}$ in the logistic regression model as follows: $s_\theta(\mathbf{X}) = \left( 1 + e^{\theta_0 + \theta_1 X_1 + \theta_2 X_1^2 + \ldots + \theta_{2d-1} X_d + \theta_{2d} X_d^2 }  \right)^{-1}$. The model parameter vector $\theta$ is estimated using least-square fitting as in \cite{chen2017oracle}.}

{\color{black} To conduct simulation experiments, we use essentially the same setup as in Cannemela example in the previous subsection. That is, we determine the failure threshold $l$ using a Monte Carlo simulation (with $10^7$ replications). We use the same simulation budget for both methods: the metamodel-based SIS uses the pilot sample of size 3000 and the IS sample of size 10,000; the CE-SIS method uses $n^{(0)} = 3000$ and $n^{(t)} = 1000$ for $t=1,\ldots,10$. For the pilot/initial sample of both methods, we generate $\mathbf{X}$ from the uniform distribution over the $d$-dimensional hypercube $(-5,5)^d$. The simulation experiment is repeated 500 times to obtain the results. Using this example, we study performances of the CE-SIS algorithm with respect to the following two aspects:
\begin{itemize}
    \item Dimension $d$ of $\mathbf{X}$: $1, 2, 4$ in Table~\ref{tab:ex2-high}.
    \item Standard deviation $\sigma$ of $Y|\mathbf{X}$: $1, 2, 4$ in Table~\ref{tab:ex2-high}.
\end{itemize}}


{\color{black} Table~\ref{tab:ex2-high} shows, as expected, all SIS methods perform worse as the dimension $d$ of $\mathbf{X}$ and the standard deviation $\sigma$ of $Y|\mathbf{X}$ get larger. These results echo the existing knowledge from the literature. Specifically, it is known that when $d$ increases, IS suffers as it becomes difficult to narrow down the important region \cite{au2003}. Also, when $\sigma$ is larger, it is known that the greater randomness within the stochastic simulation model makes SIS perform worse \cite{Choe2015}. An interesting finding is that CE-SIS performs better than the metamodel-based SIS when $d$ and $\sigma$ are small. But, as $d$ and $\sigma$ get larger, inferring the data-generating structure from data becomes harder for the CE-SIS algorithm, so that the metamodel-based SIS performs better as it uses the partial knowledge of the data-generating structure (\textit{i.e.}, symmetry with respect to the origin).}    

\begin{table}[!htbp]
\centering
{\color{black}\caption{Comparison between CE-SIS, metamodel-based SIS, and optimal SIS with respect to different dimension $d$ of $\mathbf{X}$ and standard deviation $\sigma$ of $Y|\mathbf{X}$. The target probability $\mathbb{P}\left( Y >l \right)$ to estimate is fixed at 0.01. $\frac{m}{n}$ ratio is fixed at 0.3. The results are based on 500 experiments.}}
\label{tab:ex2-high}
{\color{black}\begin{tabular}{cccccc}
\hline
Parameter                           & Threshold $l$ &Method     & Mean    & S.E. & CMC Ratio \\ \hline\hline
$d = 1, \sigma = 1$                 & 10.27   & CE-SIS        & 0.009940&	0.000334      & 14.65\%      \\
                                         &   & Metamodel     & 0.009928&	0.000523        & 35.96\%      \\
                                         &   & Optimal SIS     &0.009986&	0.000215        & 6.06\%      \\                 
$d = 1, \sigma = 2$                 & 11.43  & CE-SIS         &0.009982&	0.000664     & 57.97\%     \\ 
                                         &   & Metamodel     & 0.009983&	0.000808        & 85.65\%      \\
                                         &   & Optimal SIS     &0.009948&	0.000434        & 24.68\%      \\    
$d = 1, \sigma = 4$                 & 14.98   & CE-SIS        & 0.009998&	0.000915      & 109.82\%      \\
                                         &   & Metamodel     & 0.010041&	0.000879       & 101.56\%      \\
                                         &   & Optimal SIS     & 0.010069&	0.000762        & 76.30\%      \\    
                 \hline
$d = 2, \sigma = 1$                 & 9.37  & CE-SIS          & 0.009957&	0.000634        & 52.74\%     \\
                                         &   & Metamodel     & 0.010039&	0.000616      & 49.76\%      \\
                                         &   & Optimal SIS     & 0.009999&	0.000301        & 11.90\%      \\              
$d = 2, \sigma = 2$                 & 10.86   & CE-SIS        &  0.009967&	0.000759      &75.64\%      \\
                                         &   & Metamodel     & 0.010002&	0.000839        & 92.45\%      \\
                                         &   & Optimal SIS     & 0.009957&	0.000623        & 51.03\%      \\                  
$d = 2, \sigma = 4$                 & 14.85   & CE-SIS        & 0.010028&	0.001332     &232.84\%      \\
                                         &   & Metamodel     & 0.010054&	0.000977       & 125.38\%      \\
                                         &   & Optimal SIS     & 0.009999&	0.000833        & 91.22\%      \\                  
\hline
$d = 4, \sigma = 1$                 & 8.70 & CE-SIS           & 0.010025&	0.000750        & 73.87\%     \\
                                         &   & Metamodel     & 0.010004&	0.000705      & 65.22\%      \\
                                         &   & Optimal SIS     & 0.009992&	0.000451        & 26.65\%      \\                  
$d = 4, \sigma = 2$                 & 10.48   & CE-SIS        & 0.009839&	0.001679       &370.05\%      \\
                                         &   & Metamodel     & 0.009922&	0.000956       & 120.10\%      \\
                                         &   & Optimal SIS     & 0.009945&	0.000780        & 79.84\%      \\                   
$d = 4, \sigma = 4$                 & 14.76  & CE-SIS         & 0.009372&	0.003848      &1944.85\%      \\
                                         &   & Metamodel     & 0.010009&	0.001022        & 137.24\%      \\
                                         &   & Optimal SIS     & 0.009986&	0.000904        & 107.21\%      \\                
\hline
\end{tabular}}
\end{table}





{\color{black}We hope that the numerical studies here provide informative demonstration of the advantages and limitations of the CE-SIS method. This metamodel-\textit{free} approach generally performs well in comparison to the metamodel-based approaches that assume partial knowledge of the data-generating structure. But, a user should be careful when a) the original distribution of input $\mathbf{X}$ is heavy-tailed, b) $\mathbf{X}$ is multivariate, or c) the conditional distribution of $Y|\mathbf{X}$ is suspected to have generally large variances across $\mathbf{X}$.} 

\section{Case Study}\label{sec:case_study}

This section presents an application of the CE-SIS method to the case study on wind turbine reliability evaluation. We compare the performance between the CE-SIS method and the metamodel-based method \cite{Choe2015}, using the same CMC ratio metric in Section~\ref{sec:num_ex}. 

The reliability of a wind turbine is subject to stochastic weather \citep{Byon2010a}. To incorporate the randomness into the reliability evaluation of a turbine design, the international standard, IEC  61400-1 \citep{IEC2005}, requires the turbine designer to use stochastic simulations. {\color{black}The reliability of a wind turbine is typically evaluated} using computationally expensive aerodynamic simulators (each simulation run takes roughly 1 minute on a typical PC available nowadays) developed by the U.S. National Renewable Energy Laboratory.  Specifically, TurbSim is used to generate a 3-dimensional time-marching wind profile, and FAST is used to simulate a turbine's structural load responses \cite{Choe2015, Jonkman2005, Jonkman2009}. Each simulation represents a 10-min simulated operation of the turbine. {\color{black}The national lab acknowledges the computational challenges in using the stochastic simulations and makes efforts to accelerate the simulation experiment \cite{graf2018adaptive,graf2016high}. We adopt the identical simulation setup used in the national lab's benchmark simulation experiment in \cite{moriarty2008}. An interested reader is referred to \cite{moriarty2008} for details of the simulation setup, as it involves long technical specifications of the wind profile and the wind turbine design.} 

\begin{figure}[!htb]
    \begin{center}
        \includegraphics[width=0.5\textwidth,trim=2.5in 1.3in 4.5in 1in, clip]{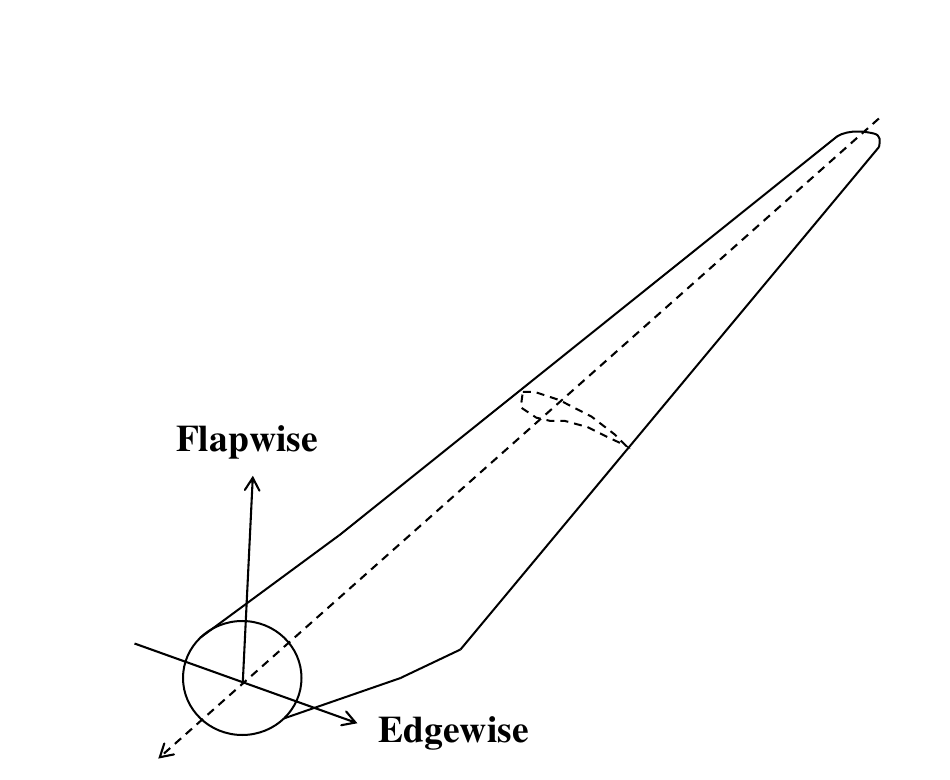}
    \caption{{\color{black}Diagram of a wind turbine blade where the directions of edgewise and flapwise bending moments are indicated at the blade root.}}
    \label{fig:turbine_blade}
    \end{center}
\end{figure}

We are interested in estimating the probability of a failure event which is defined as the bending moment at a wind turbine's blade root exceeding a specified threshold. In particular, we study two bending moment types: \textcolor{black}{edgewise bending moment (which is parallel to the blade edge) and flapwise bending moment (which is perpendicular to the blade plane) (see Figure~\ref{fig:turbine_blade})}. We estimate the probability that a bending moment exceeds a threshold $l$, where $l$ = 8,600 kNm for edgewise bending moment and $l$ = 13,800 kNm for flapwise bending moment, both of which occur with around 5\% probability. 

The input wind speed, $X$, is drawn from the truncated Rayleigh density \cite{Choe2015,choe2015extload} defined as: 
\[
f(x) = \frac{f_R(x)}{F_R(x_{out}) - F_R(x_{in})},
\]
where $f_R$ is the untruncated Raleigh density with the shape parameter, $10\sqrt{2/\pi}$, and $F_R$ is the corresponding cummulative distribution function. The cut-in and cut-out wind speed are $x_{in}$ = 3 m/s and $x_{out}$ = 25 m/s, respectively, denoting the range of wind speeds for which a wind turbine operates. 



In the metamodel-based method in \cite{Choe2015}, the conditional probability $s(X) = \mathbb{P}(Y > l \mid X)$ is approximated using a nonhomogenous Generalized Extreme Value (GEV) distribution:
\begin{equation}
    P(Y \leq y \mid X = x) =
    \begin{cases}
      \exp\left(-\left(1 + \xi\left( \frac{y-\mu(x)}{\sigma(x)}\right)\right)^{-1/\xi} \right), & \text{for}\ \xi \neq 0 \\
      \exp\left(-  \exp\left( -\frac{y-\mu(x)}{\sigma(x)}\right)\right), & \text{otherwise,} 
    \end{cases}
  \end{equation}
where $\mu(x)$ and $\sigma(x)$ are the location and scale parameter functions, respectively, modeled with cubic smoothing spline functions under the GAMLSS framework. The shape parameter $\xi$ is fixed as a constant, with estimated value $\hat \xi =$ $-0.0359$ $(-0.0529)$ for the edgewise (flapwise) case. The goodness-of-fit of the GEV distribution is tested using the Kolmogorov-Smirnov test. The metamodel is built using a pilot sample of 600 simulation replications and in addition, the metamodel-based method uses 1000 (2000) replications for the failure probability estimation for edgewise (flapwise) bending moment. 





In the CE-SIS method, we allocate the same simulation budget as the metamodel-based method. We use $n^{(0)} = 600$ and $n^{(t)} = 100$ (200) for $t=1,\ldots,10$ for the edgewise (flapwise) case. 
 
Table~\ref{tab:case} compares the results based on 50 repetitions of each method. The CE-SIS method has slightly smaller (larger) standard error than the metamodel-based method for the edgewise (flapwise) bending moment. Accordingly, both methods save the similar level of computational resource compared to CMC, as indicated by the CMC Ratio. 
In the metamodel-based method, the metamodel is carefully built by fitting a nonhomogeneous GEV distribution to the pilot data. We can see that the performance of the CE-SIS method is comparable to that of the metamodel-based SIS with a high quality  metamodel. Note that since CE-SIS is an automated method, it can be particularly helpful when building a metamodel is difficult. We also observe that the computational saving in the edgewise case is more substantial than the flapwise case. This is because the approximated SIS density is not very different from the original density in the flapwise case (this phenomenon is first observed and discussed extensively in \cite{Choe2015}, to which an interested reader is referred). Nevertheless, we see that the CE-SIS method, without requiring domain knowledge about the underlying process, can still capture this information satisfactorily. 
\begin{table}[h]
\centering
\caption{Comparison between the metamodel-based SIS and the CE-SIS method for the case study}
\label{tab:case}
\begin{tabular}{ccccc}
\hline
Response & Method    & Mean   & S.E. & CMC Ratio \\ \hline\hline
Edgewise & Metamodel & 0.0486 & 0.0018         & 7.0\%       \\
         & CE-SIS      & 0.0486 & 0.0015         & 4.9\%       \\
Flapwise & Metamodel & 0.0514 & 0.0028         & 32\%        \\
         & CE-SIS      & 0.0535 & 0.0030         & 37\%        \\ \hline
\end{tabular}
\end{table}



\vspace{-.3cm}

\section{Conclusion}
We propose a method called the cross-entropy based stochastic importance sampling (CE-SIS) which can efficiently construct an importance sampling density for a stochastic simulation model. The CE-SIS method uses an EM algorithm to minimize the estimated cross-entropy from a candidate IS density to the optimal IS density while penalizing the model complexity concurrently. The method automatically refines the estimated IS density, thus not requiring the specific domain knowledge for building a metamodel, which is often difficult or time-consuming in practice. We focus on a candidate IS density expressed as a Gaussian mixture model, which is both flexible and computationally efficient, while the extension to other mixture models is possible. The application of the cross-entropy information criterion allows a sound choice of the mixture model complexity for the CE-SIS method. By aggregating all the data from previous iterations and effectively allocating the number of replications at each input value, the CE-SIS method utilizes the available information efficiently under a given simulation budget. {\color{black}The numerical studies and case study show the advantages and limitations of the CE-SIS method in comparison to the metamodel-based SIS and crude Monte Carlo simulation.}






\clearpage
{\color{black}
\section*{Appendix 1: Approximation of $N_i$}
We seek an asymptotic approximation of the optimal allocation size, 
\begin{align*}
N_i &= n
\frac
{ 		 	\sqrt{    \frac{n \left(1-  s\left(\mathbf{X}_i\right) \right)  }{1 + \left(n-1\right)  s\left(\mathbf{X}_i\right) }     }    	   }
{ \sum_{j=1}^{m}   \sqrt{    \frac{n \left(1-  s\left(\mathbf{X}_j\right) \right)  }{1 + \left(n-1\right)  s\left(\mathbf{X}_j\right) }     }      }, \;\;\; i=1,\ldots,m. 
\end{align*}
First, we show $s\!\left(\mathbf{X}_i\right) \approx  \frac{\hat{P}_{SIS}}{w(\mathbf{X}_i)}$. For a large $n \gg \max_{i=1}^{m} {(1-s\!\left(\mathbf{X}_i\right))}/{s\!\left(\mathbf{X}_i\right)}$, we can approximate
 \begin{align*}
 q_{SIS}\!\left(\mathbf{X}_i\right)	&= \frac{1}{C_{q}} f\!\left(\mathbf{X}_i\right)	\sqrt{	\frac{1}{n} s\!\left(\mathbf{X}_i\right)\left( 1- s\!\left(\mathbf{X}_i\right)\right)    +		 s\!\left(\mathbf{X}_i\right)^2 		}     
 \\&\approx \frac{1}{C_{q}} f\!\left(\mathbf{X}_i\right)	\sqrt{	     s\!\left(\mathbf{X}_i\right)^2 		}
 \\&= \frac{1}{C_{q}} f\!\left(\mathbf{X}_i\right)	s\!\left(\mathbf{X}_i\right) 
 \end{align*}
 for any $i = 1,\ldots,m$.  This asymptotic approximation may be not good for some $N_i$ if $s\!\left(\mathbf{X}_i\right)$ is close to zero. However, in that case, $q(\mathbf{X}_i)$ is small too, and such $\mathbf{X}_i$ is unlikely to be sampled in the first place.  Therefore, we can approximate 
 \begin{align*}
 s\!\left(\mathbf{X}_i\right) &\approx C_{q} \frac{q_{SIS}\!\left(\mathbf{X}_i\right)}{f\!\left(\mathbf{X}_i\right)} \\
 &\approx \frac{C_{q}}{w(\mathbf{X}_i)}  ,
 \end{align*}
 where the second approximation is based on the fact that the estimated IS density approximates the optimal density. 
 
 Furthermore, for a large $n \gg \max_{i=1}^{m} {(1-s\!\left(\mathbf{X}_i\right))}/{s\!\left(\mathbf{X}_i\right)}$, we can also approximate
 \begin{align*}
 C_{q} &= \int_{\mathcal{X}_f} f\!\left(\mathbf{x}\right)	\sqrt{	\frac{1}{n}s\!\left(\mathbf{x}\right)\cdot\left( 1-s\!\left(\mathbf{x}\right)\right)    +		 s\!\left(\mathbf{x}\right)^2 } \,\mathrm{d}\mathbf{x} 
  \\&\approx  \int_{\left\lbrace \mathbf{x} :\; s\left(\mathbf{x}\right) > 1/(n+1)\right\rbrace } f\!\left(\mathbf{x}\right)	\sqrt{	\frac{1}{n}s\!\left(\mathbf{x}\right)\cdot\left( 1-s\!\left(\mathbf{x}\right)\right)    +		 s\!\left(\mathbf{x}\right)^2 }  \,\mathrm{d}\mathbf{x} 
 \\&\approx  \int_{\left\lbrace \mathbf{x} :\; s\left(\mathbf{x}\right) > 1/(n+1)\right\rbrace } f\!\left(\mathbf{x}\right)	s\!\left(\mathbf{x}\right) \,\mathrm{d}\mathbf{x} 
 \\&\approx\hat{P}_{SIS} ,
 \end{align*}
 where $\mathcal{X}_f$ is the support of $f$.
 Thus, it follows that 
 \begin{align*}
 s\!\left(\mathbf{X}_i\right) &\approx  \frac{\hat{P}_{SIS}}{w(\mathbf{X}_i)} .
 \end{align*}
 Therefore, for a large $n$,
 \begin{align*}
 N_i &\propto  \sqrt{    \frac{n \left(1-  s\left(\mathbf{X}_i\right) \right)  }{1 + \left(n-1\right)  s\left(\mathbf{X}_i\right) }     } 
 \\&\approx  \sqrt{    \frac{ 1-  s\left(\mathbf{X}_i\right) }{s\left(\mathbf{X}_i\right)   }     } 
 \\&\approx \sqrt{    \frac{ 1-  \frac{\hat{P}_{SIS}}{w(\mathbf{X}_i)} }{\frac{\hat{P}_{SIS}}{w(\mathbf{X}_i)}   }     }
 \\&\propto  \sqrt{w(\mathbf{X}_i)-\hat{P}_{SIS}} .
 \end{align*}
 Although it does not happen frequently, if $w(\mathbf{X}_i)-\hat{P}_{SIS} \leq 0$, then we set the corresponding $N_i$ as $1$, the smallest allocation possible to maintain the unbiasedness of the SIS estimator.

}



\section*{Appendix 2: Implementation details of the CE-SIS algorithm}
Since the EM algorithm will lead to a local optimum for non-convex problems, we use 10 random initializations of $\boldsymbol{\theta}$ and choose the best minimizer $\boldsymbol{\hat\theta}$ of $\bar{\mathcal{C}}^{(t-1)}_{SIS}(\boldsymbol{\hat\theta})$ in \eqref{eq:CE_SIS_est_hat_theta} to reduce the impact of initial guess of $\boldsymbol{\theta}$ on the algorithm's performance \citep{figueiredo2002}. 

By monitoring the condition numbers of the Gaussian components' covariances \citep{figueiredo2002}, the number of components $k$ can also be controlled to prevent over-fitting issue within the EM algorithm. We choose a singularity threshold of $10^5$ for the covariance matrix. Exceeding the threshold will signify an ill-conditioned matrix. If more than a half of the 10 initializations observe ill-conditions, we stop the EM algorithm and consider $k_{\max}$ is reached for that iteration. 

Checking the convergence of the EM algorithm is through monitoring the reduction of $\bar{\mathcal{C}}^{(t-1)}_{SIS}(\boldsymbol{\hat\theta})$ in \eqref{eq:CE_SIS_est_hat_theta}. In the numerical studies in Section~\ref{sec:num_ex}, iterating the updating equations in the EM algorithm is stopped if the reduction of $\bar{\mathcal{C}}^{(t-1)}_{SIS}(\boldsymbol{\hat\theta})$ is less than 1\% or a specified maximum number of iterations is reached.

{\color{black}For the $N_i^{(t)}$ allocation at each iteration $t$, we round the calculated $N_i^{(t)}$ to the nearest integer and we set $N_i^{(t)} = 1$ if the rounding leads to $0$ (see Algorithm~\ref{alg:alg1}). The positive or negative difference between $\sum_{i=1}^{m} N_i^{(t)}$ and $n^{(t)}$ due to the rounding is distributed across the $N_i^{(t)}$'s sequentially to make sure we strictly stay within the simulation budget at each iteration. Note that SIS is generally insensitive to variations of $N_i^{(t)}$'s, as shown in Table~\ref{tab:ex2-ratio} and reported in \cite{Choe2015}.}  

{\color{black}\section*{Acknowledgements}
This work was partially supported by the National Science Foundation (NSF grant CMMI-1824681).}

\section*{References}

\bibliography{mybibfile}

\end{document}